\documentclass[11pt]{article}
\usepackage{axodraw}
\usepackage{epsfig}
\newcommand{\la}[1]{\label{#1}}
\newcommand{\be}{\begin{equation}}
\newcommand{\ee}{\end{equation}}
\newcommand{\ba}{\begin{eqnarray}}
\newcommand{\ea}{\end{eqnarray}}
\newcommand{\bi}{\begin{itemize}}
\newcommand{\ei}{\end{itemize}}
\newcommand{\rmi}[1]{{\mbox{\scriptsize #1}}}
\newcommand{\nr}[1]{(\ref{#1})}
\newcommand{\tr}{{\rm Tr\,}}
\newcommand{\re}{\mathop{\rm Re}}
\newcommand{\Hc}{{\rm H.c.\ }}

\newcommand{\nn}{\nonumber \\}
\newcommand{\fr}[2]{{\frac{#1}{#2}}}

\renewcommand{\vec}[1]{{\bf #1}}

\newcommand{\trp}{{\rm Tr}_v\,}
\newcommand{\str}{{\rm Str\,}}
\renewcommand{\a}{r}    %{\alpha}
\renewcommand{\b}{s}    %{\beta}
\renewcommand{\c}{u}    %{\gamma}
\renewcommand{\d}{v}    %{\delta}
\renewcommand{\k}{k}    %{\kappa}
\renewcommand{\l}{l}    %{\lambda}
\newcommand{\ta}{\tilde r}     %{\tilde{\a}}
\newcommand{\tb}{\tilde s}     %{\tilde{\b}}
\newcommand{\tc}{\tilde u}     %{\tilde{\c}}
\newcommand{\td}{\tilde v}     %{\tilde{\d}}
\newcommand{\1}{d} 
\newcommand{\2}{u} 
\newcommand{\3}{s} 
\newcommand{\4}{c} 
\newcommand{\RR}{{\rm I\kern -.2em  R}}
\newcommand{\eq}{Eq.~}
\newcommand{\eqs}{Eqs.~}
\newcommand{\fig}{Fig.~}

\newcommand{\se}{Sec.~}

\def\lsi{\raise0.3ex\hbox{$<$\kern-0.75em\raise-1.1ex\hbox{$\sim$}}}
\def\gsi{\raise0.3ex\hbox{$>$\kern-0.75em\raise-1.1ex\hbox{$\sim$}}}
\newcommand{\lsim}{\mathop{\lsi}}

\makeatletter \@addtoreset{equation}{section} \makeatother
\renewcommand{\theequation}{\arabic{section}.\arabic{equation}}
\makeatletter
\renewcommand\section{\@startsection {section}{1}{\z@}%
                                   {-5.5ex \@plus -1ex \@minus -.2ex}% bfr-skip
                                   {2.3ex \@plus.2ex}%
                                   {\normalfont\large\bfseries}}
\renewcommand\subsection{\@startsection{subsection}{2}{\z@}%
                                     {-3.25ex\@plus -1ex \@minus -.2ex}%
                                     {1.5ex \@plus .2ex}%
                                     {\normalfont\normalsize\bfseries}}
\renewcommand\thesection {\@arabic\c@section}
\renewcommand\thesubsection   {\thesection.\@arabic\c@subsection}
\renewcommand{\@seccntformat}[1]{%
\csname the#1\endcsname.\hspace{1.0em}}
\makeatother
\newcommand{\picc}[1]{\;\parbox[c]{60pt}{\begin{picture}(60,30)(0,0)
\SetWidth{1.0}\SetScale{1.0} #1 \end{picture}}\;}
\newcommand{\piccb}[1]{\;\parbox[c]{75pt}{\begin{picture}(75,30)(0,0)
\SetWidth{1.0}\SetScale{1.0} #1 \end{picture}}\;}
\newcommand{\piccc}[1]{\;\parbox[c]{90pt}{\begin{picture}(90,30)(0,0)
\SetWidth{1.0}\SetScale{1.0} #1 \end{picture}}\;}

\def\Asc(#1,#2)(#3,#4,#5){\CArc(#1,#2)(#3,#4,#5)}
\def\Lsc(#1,#2)(#3,#4){\Line(#1,#2)(#3,#4)}
\def\DAsc(#1,#2)(#3,#4,#5){\DashCArc(#1,#2)(#3,#4,#5){3}}
\def\DLsc(#1,#2)(#3,#4){\DashLine(#1,#2)(#3,#4){3}}
\def\TAsc(#1,#2)(#3,#4,#5){\SetWidth{2.0}\CArc(#1,#2)(#3,#4,#5)\SetWidth{1.0}}
\def\TLsc(#1,#2)(#3,#4){\SetWidth{2.0}\Line(#1,#2)(#3,#4)\SetWidth{1.0}}
\def\TDAsc(#1,#2)(#3,#4,#5){\SetWidth{2.0}\DashCArc(#1,#2)(#3,#4,#5){3}%
\SetWidth{1.0}}
\def\TDLsc(#1,#2)(#3,#4){\SetWidth{2.0}\DashLine(#1,#2)(#3,#4){3}%
\SetWidth{1.0}}
\def\CTopomeas(#1,#2){\picc{#1(0,15)(20,15) #2(20,15)(40,15)% 
\SetWidth{2.0} \Line(17,12)(23,18) \Line(17,18)(23,12) \SetWidth{1.0}%
\GBoxc(0,15)(5,5){1} \GBoxc(40,15)(5,5){1} }}
\def\CTopomass(#1,#2){\picc{#1(0,15)(20,15) #2(20,15)(40,15)% 
\GCirc(20,15){3}{0}% 
\GBoxc(0,15)(5,5){1} \GBoxc(40,15)(5,5){1} }}
\def\CTopoL4(#1,#2){\picc{#1(0,15)(20,15) #2(20,15)(40,15)% 
\GBoxc(20,15)(5,5){0}% 
\GBoxc(0,15)(5,5){1} \GBoxc(40,15)(5,5){1}%
\Text(20,25)[c]{$L_4$} }}
\def\CTopoin(#1,#2,#3){\picc{#1(0,15)(20,15) #2(20,15)(40,15)% 
#3(20,22)(7,0,360)% 
\GBoxc(0,15)(5,5){1} \GBoxc(40,15)(5,5){1} }}
\def\CTopomassin(#1,#2,#3){\picc{#1(0,15)(20,15) #2(20,15)(40,15)% 
#3(20,22)(7,0,360)% 
\GCirc(20,15){3}{0}% 
\GBoxc(0,15)(5,5){1} \GBoxc(40,15)(5,5){1} }}
\def\CTopocucu(#1,#2){\picc{#1(20,-5)(28,45,135) #2(20,35)(28,225,315)%
\GBoxc(0,15)(5,5){1} \GBoxc(40,15)(5,5){1} }}
\def\CTopocu(#1,#2){\picc{#1(0,22)(7,0,360) #2(0,15)(40,15)%
\GBoxc(0,15)(5,5){1} \GBoxc(40,15)(5,5){1} }}
\def\Topomeas(#1,#2,#3){\piccc{#1(0,15)(20,15) #2(20,15)(40,15)% 
#3(40,15)(80,15) %
\SetWidth{2.0} \Line(17,12)(23,18) \Line(17,18)(23,12) \SetWidth{1.0}%
\GBoxc(0,15)(5,5){1} \GBoxc(80,15)(5,5){1} \GCirc(40,15){3}{1} }}
\def\Topomass(#1,#2,#3){\piccc{#1(0,15)(20,15) #2(20,15)(40,15)% 
#3(40,15)(80,15) \GCirc(20,15){3}{0}% 
\GBoxc(0,15)(5,5){1} \GBoxc(80,15)(5,5){1} \GCirc(40,15){3}{1} }}
\def\TopoL4(#1,#2,#3){\piccc{#1(0,15)(20,15) #2(20,15)(40,15)% 
#3(40,15)(80,15) \GBoxc(20,15)(5,5){0}% 
\GBoxc(0,15)(5,5){1} \GBoxc(80,15)(5,5){1} \GCirc(40,15){3}{1}%
\Text(20,25)[c]{$L_4$} }}
\def\Topoin(#1,#2,#3,#4){\piccc{#1(0,15)(20,15) #2(20,15)(40,15)% 
#3(40,15)(80,15) #4(20,22)(7,0,360)% 
\GBoxc(0,15)(5,5){1} \GBoxc(80,15)(5,5){1} \GCirc(40,15){3}{1} }}
\def\Topomassin(#1,#2,#3,#4){\piccc{#1(0,15)(20,15) #2(20,15)(40,15)% 
#3(40,15)(80,15) #4(20,22)(7,0,360) \GCirc(20,15){3}{0}% 
\GBoxc(0,15)(5,5){1} \GBoxc(80,15)(5,5){1} \GCirc(40,15){3}{1} }}
\def\Topoinop(#1,#2,#3){\piccc{#1(0,15)(38,8) #2(42,8)(80,15)% 
#3(40,15)(7,0,360)%
\GBoxc(0,15)(5,5){1} \GBoxc(80,15)(5,5){1} \GCirc(40,22){3}{1} }}
\def\Topomassinop(#1,#2,#3){\piccc{#1(0,15)(38,8) #2(42,8)(80,15)% 
#3(40,15)(7,0,360) \GCirc(40,8){3}{0}%
\GBoxc(0,15)(5,5){1} \GBoxc(80,15)(5,5){1} \GCirc(40,22){3}{1} }}
\def\Topocuop(#1,#2,#3){\piccc{#1(20,-5)(28,45,135) #2(20,35)(28,225,315)%
#3(40,15)(80,15)%
\GBoxc(0,15)(5,5){1} \GBoxc(80,15)(5,5){1} \GCirc(40,15){3}{1}}}
\def\Topocucu(#1,#2,#3){\piccc{#1(40,-25)(56,45,135) #2(0,15)(40,15)%
#3(40,15)(80,15)%
\GBoxc(0,15)(5,5){1} \GBoxc(80,15)(5,5){1} \GCirc(40,15){3}{1}}}
\def\Topocu(#1,#2,#3){\piccc{#1(0,22)(7,0,360) #2(0,15)(40,15)%
#3(40,15)(80,15)%
\GBoxc(0,15)(5,5){1} \GBoxc(80,15)(5,5){1} \GCirc(40,15){3}{1}}}
\def\Topocucuop(#1,#2,#3){\piccc{#1(20,-5)(28,45,135) #2(20,35)(28,225,315)%
#3(40,-25)(56,45,135)%
\GBoxc(0,15)(5,5){1} \GBoxc(80,15)(5,5){1} \GCirc(40,15){3}{1}}}
\def\Topoop(#1,#2,#3){\piccc{#2(0,15)(40,15) #3(40,15)(80,15)%
#1(40,22)(7,0,360)%
\GBoxc(0,15)(5,5){1} \GBoxc(80,15)(5,5){1} \GCirc(40,15){3}{1}}}
\def\Defmeas(#1,#2){\piccb{#1(0,15)(15,15) #2(15,15)(30,15)%
\SetWidth{2.0} \Line(12,12)(18,18) \Line(12,18)(18,12) \SetWidth{1.0}
\Text(30,15)[l]{ $=$ measure,} }}
\def\Defmass(#1,#2){\piccb{#1(0,15)(15,15) #2(15,15)(30,15)%
\GCirc(15,15){3}{0} \Text(30,15)[l]{ $=$ mass} }}
\def\Defmassin(#1,#2,#3,#4){\piccb{#1(0,15)(15,15) #2(15,15)(30,15)%
#3(15,15)(15,25) #4(15,15)(15,5)%
\GCirc(15,15){3}{0} \Text(30,15)[l]{ $=$ mass} }}
\def\Defcurr(#1){\picc{#1(0,15)(15,15) \GBoxc(0,15)(5,5){1} 
\Text(15,15)[l]{ $=$ current,} }}
\def\Defoper(#1,#2){\piccb{#1(0,15)(15,15) #2(15,15)(30,15)%
\GCirc(15,15){3}{1} \Text(30,15)[l]{ $=$ operator,} }}
\def\Deffield(#1,#2){\piccb{#1(0,15)(20,15)%
\Text(20,15)[l]{ $=$ #2,} }}
\begin{document}

\begin{titlepage}
\begin{flushright}
CERN-TH/2002-356\\
FTUV-02-1210\\
IFIC/02-59\\
hep-lat/0212014\\
\end{flushright}
\begin{centering}
\vfill

\mbox{\Large\bf Correlators of left charges and weak operators} 

\vspace*{0.1cm}

\mbox{\Large\bf in finite volume chiral perturbation theory}

\vspace*{0.8cm}

P. Hern\'andez\footnote{On leave
from Dept. de F\'{\i}sica Te\'orica, Universidad de Valencia.}  
and
M. Laine

\vspace*{0.8cm}

Theory Division, CERN, CH-1211 Geneva 23, Switzerland

\vspace*{0.8cm}

{\bf Abstract}
 
\end{centering}
 
\vspace*{0.4cm}

\noindent
We compute the two-point correlator 
between left-handed flavour charges, and the three-point correlator 
between two left-handed charges and one strangeness 
violating  $\Delta I=3/2$
weak operator, at next-to-leading order in finite volume
SU(3)$_L \times$SU(3)$_R$ chiral perturbation theory, 
in the so-called $\epsilon$-regime. Matching these 
results with the corresponding lattice measurements would in principle
allow to extract the pion decay constant $F$, and the effective
chiral theory parameter $g_{27}$, 
which determines the $\Delta I = 3/2$ amplitude
of the weak decays $K\to \pi\pi$ as well as the kaon mixing
parameter $\hat B_K$ in the chiral limit. 
We repeat the calculations in the replica formulation 
of quenched chiral perturbation theory,
finding only mild modifications. In particular, 
a properly chosen ratio of the three-point and two-point functions 
is shown to be identical in the full and quenched theories at this order.
\vfill

%\noindent
%PACS numbers: 

%11.15.Ha, %        Lattice gauge theory
%11.30.Hv, %        Flavour symmetries
%11.30.Rd, %        Chiral symmetries
%12.38.Gc, %        Lattice QCD calculations
%12.39.Fe, %        Chiral Lagrangians
%\\
%Keywords:

\vspace*{1cm}
 
\noindent
%%December 2002  %% \today
January 2003

\vfill

\end{titlepage}

%%%%%%%%%%%%%%%%%%%%%%%%%%%%%%%%%%%%%%%%%%%%%%%%%%%%%%%%%%%%%%%%%%%%%%%%%%
%%%%%%%%%%%%%%%%%%%%%%%%%% SECTION %%%%%%%%%%%%%%%%%%%%%%%%%%%%%%%%%%%%%%%

\section{Introduction}

The study of the weak matrix elements involved in kaon 
physics is a long-standing topic of lattice QCD~\cite{oldies,bdspw}. 
It is however a very difficult problem for all lattice formulations 
which break the chiral symmetry explicitly, such as Wilson fermions. 
We expect that significant progress can be achieved with the new 
Ginsparg-Wilson formulations of lattice fermions~\cite{gw}--\cite{kn}, 
which possess an exact chiral symmetry in the limit of vanishing quark 
masses, resulting in an enormous simplification in weak operator mixing 
and renormalization~\cite{ha}, \cite{gwren}--\cite{algo}. 

One advantage of approaching the regime of vanishing quark masses is 
obviously that the uncertainty induced by chiral extrapolations is avoided.  
On the other hand, 
since the chiral limit implies that light mesons become massless, 
it necessarily brings with it large finite volume effects. This apparent
difficulty can, however, be turned into a useful tool: if the finite 
volume effects can be resolved analytically in terms of the infinite 
volume properties of the theory, then the infinite volume properties 
can be extracted by monitoring the volume dependence. 
The first proposal to apply finite-size scaling techniques to 
the weak $K\to \pi\pi$ amplitudes was presented in~\cite{ll}. 

A practical realisation for the finite-volume philosophy
mentioned is offered by    
Chiral Perturbation Theory ($\chi$PT). As the quark masses get smaller, 
the chiral expansion becomes more and more accurate in describing 
the dynamics of the low momentum modes of QCD (below a few hundred MeV). 
The chiral expansion in this regime is slightly more complicated than 
in infinite volume, because it requires the resummation of pion zero 
mode contributions. Gasser and Leutwyler~\cite{gl} have presented 
a systematic procedure for doing this, the so-called
$\epsilon$-expansion (see also~\cite{N}). Several observables, 
such as the quark condensate and the scalar and vector two-point 
functions, have already been computed at next-to-leading order 
in the $\epsilon$-expansion~\cite{gl}--\cite{hl2}. These quantities 
depend on the (infinite volume) chiral theory parameters, the volume, 
and the quark masses, in a way that a comparison with lattice data 
for different volumes and quark masses, allows in principle 
the extraction of the corresponding infinite volume 
low-energy couplings of $\chi$PT.

Obviously the chiral model can be extended to include 
the $|\Delta S|=1$ weak Hamiltonian~\cite{cronin}. This introduces 
a new set of low-energy constants, from which the 
physical amplitudes in kaon decays can be extracted, by working up 
to some desired order in the chiral expansion.
The determination of these low-energy constants by matching 
the matrix elements of weak operators computed in lattice QCD to 
the same observables computed in SU(3)$_L \times $SU(3)$_R$ 
$\chi$PT was proposed a long time ago~\cite{bdspw}. Calculations 
along these lines~(for a recent review, see~\cite{ni}) 
have shown that if the matching is performed 
at relative large quark masses, there are large uncertainties induced 
by the chiral extrapolations~\cite{panel}, due to the fact that 
next-to-leading order corrections involve a large number of new 
unknown couplings~\cite{p4}. Although strategies have been 
proposed~\cite{lmpsv} to measure the relevant new couplings, together 
with the leading order ones, by matching the matrix elements at several 
kinematical conditions, this is clearly a very challenging procedure, 
particularly for the $\Delta I=1/2$ kaon decays~\cite{lmpsv,gp}. 

The approach that we consider here is instead to perform 
the matching in a finite volume but close to the chiral limit, 
in the $\epsilon$-regime. The predictions of 
$\chi$PT for the weak matrix elements in terms of the low-energy couplings 
are then not the same as in infinite volume. 
We compute the correlators of 
two left-handed flavour currents, as well as the matrix elements 
of the $|\Delta S| = 1$ (or $|\Delta S| = 2$)
weak Hamiltonian with two such currents, 
at next-to-leading order in the 
$\epsilon$-expansion.  Some motivations for this approach
have been discussed in~\cite{algo}. 
The simplification brought in by approaching the 
chiral limit is manifest in the fact that at next-to-leading order none
of the unknown couplings present in the usual $p$-expansion~\cite{p4} 
contributes. We expect therefore that the determination of the leading 
order couplings should in principle be more straightforward. 
Only the 27-plet low-energy coupling contributing to $\Delta I=3/2$
kaon decays 
is considered here, due to a number of subtleties with the octet
operator~\cite{lmpsv,gp}, which will be considered elsewhere. 

The structure of the paper is as follows. 
In~\se\ref{se:weak}, we discuss the chiral model 
representation of the weak Hamiltonian at low 
energies in the SU(3)$_L \times $SU(3)$_R$ symmetric case. 
In~\se\ref{se:finvol}, we review the $\epsilon$-expansion 
of Gasser and Leutwyler for this model, and in~\se\ref{se:results}
present the results of our next-to-leading order calculations. 
Finally, in~\se\ref{se:quench}, we present the same results 
in the quenched theory. We conclude in~\se\ref{se:concl}. 

%%%%%%%%%%%%%%%%%%%%%%%%%% SECTION %%%%%%%%%%%%%%%%%%%%%%%%%%%%%%%%%%%%%%%

\section{Weak operators in the SU(3) chiral theory}
\la{se:weak}

Ignoring weak interactions, 
the QCD chiral Lagrangian possesses an SU(3)$_L\times$SU(3)$_R$ symmetry, 
broken ``softly'' by the mass terms. The Euclidean Lagrangian 
can to leading order in a momentum expansion be written as 
\be
 {\cal L}_E = 
 \frac{F_{}^2}{4} \tr \Bigl[\partial_\mu U\partial_\mu U^\dagger \Bigr] - 
 \frac{\Sigma}{2} \tr \Bigl[ 
 U M e^{i\theta/N_f} + M^\dagger U^\dagger e^{-i\theta/N_f}
 \Bigr]. 
 \la{LE}
\ee
Here $U\in $ SU(3), $\theta$ is the vacuum angle, $N_f=3$, $M$
is the quark mass matrix and, 
to leading order in the chiral expansion,  $F_{}$, $\Sigma$,  
equal the pseudoscalar decay constant and 
the chiral condensate, respectively. 
We shall for convenience take $M$ to be real and diagonal.

Weak interactions break explicitly
the SU(3)$_L\times$SU(3)$_R$ symmetry of \eq\nr{LE}. In the fundamental 
theory, the strangeness violating interactions 
responsible for kaon decays can be accurately described  
through an operator product expansion in the inverse W-boson mass.
In the CP conserving case of two generations, the effective 
weak Hamiltonian to leading order in the QCD coupling constant 
is then (see, e.g.,~\cite{hg,revs})
\be
 %% {L}_E^{(\Delta S = \pm 1)} = 
 H_w = 
 2 \sqrt{2} G_F V_{ud} V^*_{us}
 \Bigl[ 
 (\bar s \gamma_\mu P_{-} u) (\bar u \gamma_\mu P_{-} d) - 
 (\bar s \gamma_\mu P_{-} c) (\bar c \gamma_\mu P_{-} d) 
 \Bigr] + \Hc \;,
 \la{Lw_QCD}
\ee
where $G_F$ is the Fermi constant,  $V_{ij}$ 
are elements of the CKM-matrix, and $P_{\pm} = (1\pm \gamma_5)/2$. 
Taking into account QCD radiative corrections, 
the coefficients get modified, but $H_w$ can  
still be written as~\cite{wilson,cg} 
\be
 H_w = 
 2 \sqrt{2} G_F V_{ud} V^*_{us} 
 \biggl\{
 \sum_{\sigma = \pm 1}
 h_w^\sigma 
 \Bigl( [{O_{w}}]^\sigma_{suud} - [{O_{w}}]^\sigma_{sccd}\Bigr) 
 + h_m 
 [{O_{m}}]_{sd} \biggr\}
 + \Hc \;, \la{Hw}
\ee
where $h_{w}^\pm, h_{m}$ are regularisation dependent
dimensionless Wilson coefficients, and we have introduced the notation 
\ba
 [ O_{w} ]^\sigma_{\a\b\c\d} & \equiv & 
 \fr12 
 \Bigl( [ O_{w} ]_{\a\b\c\d} + 
 \sigma [ O_{w} ]_{\a\b\d\c} \Bigr)\;, \la{Owplus} \\
 {[ O_{w} ]}_{\a\b\c\d} & \equiv & 
 (\bar\psi_{\a} \gamma_\mu P_{-} \psi_{\c}) 
 (\bar\psi_{\b} \gamma_\mu P_{-} \psi_{\d}) \;,  
 \la{O_QCD} \\
 {[ O_{m} ]}_{sd} ~~ & \equiv & 
 (m_c^2 - m_u^2) \{ [\bar\psi M ]_{\3} P_- \psi_{\1} + 
 \bar\psi_{\3} P_+ [M^\dagger \psi]_{\1} \} \;.
 \la{Lw_QCD_general} \la{O2_QCD}
\ea
Here $r,s,u,v$ are generic flavour indices, 
while $u,d,s,c$ denote the physical flavours.
The Wilson coefficients have been computed
also for Ginsparg-Wilson ``overlap'' fermions~\cite{cg},
apart from $h_{m}$, which remains undetermined. 
According to \eq\nr{Lw_QCD}, the leading order values are
$h_{w}^\pm = 1, h_{m}=0$.

In order to match the Hamiltonian
of~\eq\nr{Hw} to the one in the SU(3) chiral theory, 
the first step is to decompose it into irreducible representations of 
the SU(3)$_L\times$SU(3)$_R$ flavour group, 
present at low energies. 
For completeness, we review
the general formulae for the decomposition in Appendix A. 
The weak operators are singlets under SU(3)$_R$, 
and denoting  
projected operators transforming under representations
of SU(3)$_L$ with 
dimensions 27, 8 by $[\, \hat {\! O}_{w}]^+_{\a\b\c\d}$, 
$[R_{w}]^\sigma_{\a\c}$,
respectively, the weak Hamiltonian can be rewritten as 
\ba
 {H}_w & = &   
  2 \sqrt{2} G_F V_{ud} V^*_{us} 
 \biggl\{
  h_w^+ [{\hat {O}_w}]^+_{\3\2\2\1} 
  + \fr15 h_w^+ [R_w]^+_{\3\1} - h_w^- [R_w]^-_{\3\1} \nn
  & - & \fr12 (h_w^+ + h_w^-) [O_w]_{\3\4\4\1} 
  - \fr12 (h_w^+ - h_w^-) [O_w]_{\3\4\1\4} 
  + h_m [O_m]_{\3\1}
 \biggr\} + \Hc \;, 
 \la{Lw_QCD_su3}
\ea
where 
\ba
 [\hat {O}_w]^+_{\3\2\2\1} & \equiv & 
 \frac{1}{2} \Bigl\{ [O_w]_{\3\2\2\1} + [O_w]_{\3\2\1\2} 
 - \fr15 \sum_{k=u,d,s}
 \Bigl( [O_w]_{\3\k\1\k} + [O_w]_{\3\k\k\1}  \Bigr)\Bigr\} \;, 
 \la{preO27} \\
 {[R_w]}_{\3\1}^{\pm} & \equiv & 
 \fr12 \sum_{k=u,d,s} \Bigl( [O_w]_{\3\k\1\k} \pm [O_w]_{\3\k\k\1} \Bigr) \;.
 \la{preO8}
\ea
The first operator in \eq\nr{Lw_QCD_su3} transforms
under the 27-plet of the SU(3)$_L$ subgroup: it is symmetric under 
the interchange of quark or antiquark indices, and traceless. 
The remaining ones, transforming as ${\bf 3^* \otimes 3}$
and being traceless, belong to irreducible representations 
of dimension 8. 

The next step is to find the chiral analogue for this weak Hamiltonian, 
as well as for the left-handed flavour currents, which we will use
as external probes. In a convention for fermion fields where the 
Euclidean Lagrangian reads ($\gamma_\mu^\dagger = \gamma_\mu$, 
$\{\gamma_\mu,\gamma_\nu\} = 2 \delta_{\mu\nu}$)
\be
 {L}_E = \bar \psi (\gamma_\mu D_\mu + M P_- + M^\dagger P_+)\psi, 
\ee
we may define a left-handed current as 
\be
 \Bigl(J^a_\mu\Bigr)_\rmi{QCD}  
 \equiv 
 i \bar \psi_{\a}  T^a_{\a\c} \gamma_\mu P_{-} \psi_{\c}, \la{J_QCD} 
\ee
where the $T^a$ are Hermitean generators of the flavour SU(3)\footnote{%
 In fact, the only property of $T^a$ we need to assume
 is their tracelessness.}.
As usual, it is convenient to introduce an external left-handed 
flavour gauge field source, $A_\mu^a$, such that 
\be
 \Bigl(J^a_\mu\Bigr)_\rmi{QCD} = \frac{\partial {L}_E}{\partial A^a_\mu}. 
\ee 
Observables including $(J^a_\mu)_\rmi{QCD}$
can then be addressed within the chiral theory
by coupling also the pion field covariantly to $A_\mu^a$, and taking 
functional derivatives with respect to it. More concretely, 
the partial derivatives of \eq\nr{LE} are promoted to covariant ones, 
\be
 \partial_\mu U \to D_\mu U \equiv [\partial_\mu + i A^a_\mu T^a] U,
\ee
 and the left-handed current is defined as 
\be
 \Bigl(J_\mu^a\Bigr)_\rmi{$\chi$PT} \equiv
 {\cal J}_\mu^a \equiv
 \left. \Bigl( \frac{\partial {\cal L}_E}{\partial A^a_\mu} \Bigr)
 \right|_{A^a_\mu = 0} 
 = -i \frac{F_{}^2}{2} T^a _{\a\c} \Bigl(
 \partial_\mu U U^\dagger \Bigr)_{\c\a},
 \la{cJmua}
\ee
up to higher order corrections. 

We can now find the chiral analogues for the building
blocks in~\eqs\nr{Owplus}--\nr{O2_QCD}.
By a comparison of \eqs\nr{O_QCD}, 
\nr{J_QCD}, the analogue of the operator in \eq\nr{O_QCD}, denoted 
in the chiral case by $[{\cal O}_{w}]_{\a\b\c\d}$, can be written as
\be
 [{\cal O}_{w}]_{\a\b\c\d} = 
 \fr14 F_{}^4 
 \Bigl(\partial_\mu U U^\dagger\Bigr)_{\c\a}
 \Bigl(\partial_\mu U U^\dagger\Bigr)_{\d\b}\;. 
 \la{O_XPT}
\ee
The operator of \eq(\ref{O_XPT}) 
is the only one with the same symmetry 
properties as its counterpart in the fundamental theory, 
at the leading order in the chiral expansion. 
Correspondingly, we can also find the chiral 
counterpart for the operator $O_m$, by employing scalar
and pseudoscalar external sources $S^a,P^a$, defined
with the substitution $M \rightarrow M + S^a T^a - i P^a T^a$, 
and taking derivatives with respect to the sources:
\be
 [{\cal O}_{m}]_{\3\1} = 
 - (m_c^2 - m_u^2) \frac{\Sigma}{2}
 (U M e^{i\theta/N_f} + M^\dagger U^\dagger e^{-i\theta/N_f})_{\1\3} \;.
\ee 

Given these building blocks, we can determine all 
the operators (at the leading order in the chiral expansion)
transforming under the 27-plet and octet of SU(3)$_L$, 
which allows then to translate the weak Hamiltonian 
of~\eq\nr{Lw_QCD_su3} to the chiral theory.
Because some contractions are 
traceless, cf.\ \eqs\nr{tless_su3}, \nr{traceless_su3}, there
are only three such operators~\cite{bdspw}. For the physical
choice of indices, we write these as
\ba
 {\cal O}_{27} & \equiv & 
 {[\; \hat {\! {\cal O}}_{w}]}^+_{\3\2\2\1} = 
 \fr35\Bigl(  
 {[ {\cal O}_{w} ]}_{\3\2\1\2} + 
 \fr23 {[  {\cal O}_{w} ]}_{\3\2\2\1}
 \Bigr), \la{formofO} \\   
 {\cal O}_{8} & \equiv & 
 {[ {\cal R}_{w} ]}_{\3\1}^+ =  \fr12 \sum_{k=u,d,s}
 {[ {\cal O}_{w} ]}_{\3\k\k\1}, \la{formofR} \\
 {\cal O}'_{8} & \equiv & 
  \fr12 F_{}^2 {\Sigma}
  ( U M  e^{i\theta/N_f}
   +M^\dagger U^\dagger e^{-i\theta/N_f} )_{\1\3} 
   \;,
\ea
where we have made use of \eqs\nr{tless_su3}, \nr{traceless_su3}
to simplify the chiral versions of~\eqs\nr{preO27}, \nr{preO8}.
Note that in the definition of ${\cal O}'_{8}$ here we have left out
the explicit mass combination $(m_c^2-m_u^2)$, which can then
appear in the coefficient of this operator; the coefficient can, 
however, also receive other contributions, due to
mixings with operators of the same symmetries. 

We can now write down the analogue of $H_w$ in~\eq\nr{Lw_QCD_su3}
in the chiral theory. We denote it by ${\cal H}_w$. 
To again define dimensionless coefficients, 
we write ${\cal H}_w$ in the form 
\be
  {\cal H}_w \equiv  2 \sqrt{2} G_F V_{ud} V^*_{us}
  \biggl\{ 
  \fr53 g_{27} {\cal O}_{27} 
  + 2 g_8 {\cal O}_8
  + 2 g_8'{\cal O}'_8
  \biggr\} + \Hc  \;,
 \la{Lw_XPT}
\ee
where $g_{27}, g_8$ and $g_8'$ are the low-energy 
constants we are interested in~\cite{lo}. 

Now, it is easy to see that the 
amplitude for $\Delta I=3/2$ decays, 
such as $K^{\pm} \rightarrow \pi^{\pm} \pi^0$, 
is directly proportional to 
$g_{27}$, while the much faster $\Delta I=1/2$ decays of
$K^0_S$ get a comparable contribution both from $g_8$ and $g_{27}$.
(The parameter $g_8'$, on the other hand, 
does not contribute to physical kaon decays~\cite{wilson,bdspw,rc}.)
More quantitatively, a leading order analysis in infinite
volume~\cite{lo}, supplemented by phenomenologically 
determined large phase shifts~\cite{gm} in the 
amplitudes, suggests the well-known values
\ba
 |g_{27}| & \approx & 0.29 \;, \la{exp_g27} \\
 |g_8|    & \approx & 5.1 \;. 
\ea
It has been argued that 1-loop corrections in the 
chiral perturbation theory are large~\cite{p4,1loop,pp}, 
and one can therefore get agreement with experimental 
data on partial decay widths even with somewhat less 
differing values of $g_8$ and $g_{27}$, but a hierarchy 
still remains. 

In the limit $N_c\to\infty$, on the other hand, one obtains~\cite{lo} 
the ``tree-level'' values deducible from the naive conversion
of~\eq\nr{Lw_QCD_su3}, with $h_w^\pm$ as in \eq\nr{Lw_QCD},
to the corresponding chiral operators of~\eqs\nr{formofO}, 
\nr{formofR}:
\be
 g_{27} = g_8 = \fr35\;. \la{phys_g}
\ee 
Clearly a large non-perturbative enhancement of $g_8$ with 
respect to the tree-level value, and some reduction
of $g_{27}$, is needed to fit the experiment.  
The final goal is to improve on the naive estimates in~\eq\nr{phys_g}, 
by determining $g_{27}$ and $g_8$ non-perturbatively in the 
SU(3)$_L\times$SU(3)$_R$ symmetric theory. As mentioned in the 
introduction, we will in this paper discuss only $g_{27}$, due 
to various subtleties in the determination of $g_8$ (particularly 
in the quenched case). 

Let us finally recall that another physical observable determined
by $g_{27}$ is the $\hat B_K$, characterising the mixing of 
$K^0, \bar K^0$, and hence determining
the mass difference of $K_S,K_L$~\cite{dgh}.
It is defined by\footnote{%
  The parameter $B_K$ is defined identically to $\hat B_K$ but without the 
  Wilson coefficient $h_{\Delta S = 2}$, whereby it is scheme 
  and scale dependent.} 
\be
 \langle \, \bar K^0 \,|\, h_{\Delta S = 2}\, [O_w]_{ssdd} \,|\, K^0\, \rangle 
 \equiv \fr43 (m_K F_K)^2 \hat B_K, \la{BK}
\ee
where $h_{\Delta S = 2}$ is the Wilson coefficient
(see, e.g.,~\cite{herni} and references therein) related to 
the operator $[O_w]_{ssdd}$, normalised to unity at tree-level. Since 
$[O_w]_{ssdd}$ is symmetric and traceless, 
it belongs to the 27-plet.
Therefore, if $h_{\Delta S = 2}$ is replaced by $h_w^+$ in 
\eq\nr{BK}, the matrix element is in the chiral limit 
(where $m_K = m_\pi = 0, F_K = F_\pi = F$) proportional to $g_{27}$:
\be
 \fr43 \hat B_K = \frac{h_{\Delta S = 2}}{h_w^+} \cdot \fr53 g_{27}  \; . 
\ee
The tree-level value is then $\hat B_K = 3/4$, but going
to next-to-leading order 
in the large-$N_c$ approach one finds a suppression
down to $\hat B_K \simeq 0.3...0.4$~\cite{drp}. This suppression
factor is very close to what would be needed for 
$g_{27}$ to go from \eq\nr{phys_g} to \eq\nr{exp_g27}.
We may thus consider it a further motivation for the lattice
study to corroborate this prediction of the large-$N_c$ approach. 
Note that near the physical point ($m_K > 0$) 
a considerably larger value is found
(for recent reviews, see~\cite{bk}).

%%%%%%%%%%%%%%%%%%%%%%%%%% SECTION %%%%%%%%%%%%%%%%%%%%%%%%%%%%%%%%%%%%%%%

\section{Chiral perturbation theory in a finite volume}  
\la{se:finvol}

In order to determine $g_{27}$,
we will consider the lattice measurement of left-current 
two-point correlation functions in the fundamental theory 
at low enough momenta~\cite{algo}. In this regime
we assume that the effective theory gives a good description,
and thus require, to first order in the weak Hamiltonian, 
\be
 %% \left. 
 \frac{\delta^2}{\delta A^a_\mu(x) \delta A^b_\nu(y)} 
 \Bigl\langle %% J^a_\mu(x) J^b_\nu(y) 
 \int_z H_w (z)  %% L_E^{\Delta S = \pm 1}(z) 
 \Bigl\rangle_\rmi{QCD}
 %% \right|_{A^a_\mu = 0} 
 = 
 %% \left. 
 \frac{\delta^2}{\delta A^a_\mu(x) \delta A^b_\nu(y)} 
 \Bigl\langle %% {\cal J}^a_\mu(x) {\cal J}^b_\nu(y) 
 \int_z {\cal H}_w (z) %% {\cal L}_E^{\Delta S = \pm 1}(z) 
 \Bigl\rangle_\rmi{$\chi$PT},
 %% \right|_{A^a_\mu = 0}
 \la{match}
\ee
where $A^a_\mu$ is set to zero after the differentiations. 
The measurement of the left-hand-side in lattice QCD allows  
to tune the effective couplings in the weak Hamiltonian  
appearing on the right-hand-side.
In general, it is convenient even to remove 
the integral $\int_z(...)$ appearing in~\eq\nr{match}, 
since matching can also be achieved before this averaging, 
as long as the currents brought down by the functional
derivatives are far enough from each other, and $H_w(z)$.  
%% such that the functional form of the three-point correlator 
%% is reproduced by the effective theory. 
In this way we avoid complications with ``contact 
terms'', arising from operators overlapping at the same
spacetime location. It is also convenient to consider 
space-averaged charges positioned at different 
times, $x_0,y_0$, rather than local currents.

We thus discuss the product of two left-handed 
charges separated 
from the weak Hamiltonian sitting at the origin, $z\equiv 0$. 
To keep the discussion 
as general as possible, we consider matrix elements of the 
``unprojected'' operator $[{\cal O}_{w}]_{\a\b\c\d}$ in~\eq\nr{O_XPT}; 
a projection to the actual 27-plet $[\hat {\cal O}_{w}]^+_{\a\b\c\d}$ 
is then carried out by the operations in~\ref{app:su3}. Thus, 
writing the expressions in a form where their QCD analogues 
are obvious\footnote{%
 At the present order, 
 these forms differ from those obtained by taking 
 functional derivatives with respect to a flavoured gauge field, 
 as in~\eq\nr{match}, 
 only regarding unimportant ``contact terms''
 $\sim \delta(x_0), \delta(y_0), \delta(x_0-y_0)$.}, 
we will be concerned with 
\ba
 {\cal C}^{ab}(x_0) & \equiv & 
 \int_{\vec x} \langle {\cal J}^a_0(x) {\cal J}^b_0(0) \rangle, \la{Cab} \\
 {[{\cal C}_{w}]}^{ab}_{\a\b\c\d}(x_0,y_0) & \equiv & 
 \int_{\vec x} \int_{\vec y} \langle {\cal J}^a_0(x) 
 [{\cal O}_{w}]_{\a\b\c\d}(0) {\cal J}^b_0(y) \rangle, \la{Cab_abcd} 
\ea
where $\int_{\vec{x}} = \int {\rm d}^3 x$.
The computations are carried out with a finite phase $\theta$
as in \eq\nr{LE}, allowing to make predictions for the case
of a fixed topology, as well (see \se\ref{se:top}). 

Due to the vicinity of the chiral limit, we take the
volume to be a finite periodic box, of size $V = L_0L_1L_2L_3$. 
Momenta are then quantised, 
\be
 p_\mu = \frac{2\pi}{L_\mu} n_\mu, \quad n_\mu \in Z.
\ee
Since we want to be close to the chiral limit in the finite volume, 
the computation is organised according to the rules of the 
$\epsilon$-expansion~\cite{gl}. In the 
$\epsilon$-expansion one writes
\be
 U = \exp\Bigl({i \frac{2\xi}{F_{}}}\Bigr)\, U_0, \la{U_0} 
\ee
where $\xi$ has non-zero momentum modes only, while $U_0$ 
is a constant SU(3) matrix collecting the zero modes. The integration 
over $U_0$ has to be carried out exactly when 
$m \Sigma V \lsim {\cal O}(1)$, where $m$ is a quark mass, while 
the integration over the non-zero modes 
can be carried out perturbatively 
as long as $F_{} L \gg 1$.  
The power counting rules for the $\epsilon$-expansion are 
\be
 F_{} \sim {\cal O}(1), \quad
 \partial_\mu \sim {\cal O}(\epsilon), \quad
 L_\mu \sim {\cal O}(1/\epsilon), \quad
 \xi \sim {\cal O}(\epsilon), \quad
 m \sim {\cal O}(\epsilon^4). \la{epsexp}
\ee
Note that the quark mass counts as four powers of the momenta, 
rather than two as in the standard chiral expansion in infinite volume, 
where $m\sim M_\pi^2 \sim \partial_\mu^2$. The perturbative integrals
for the non-zero momentum modes are computed with the measure
\be
% \int_x = \int d^4 x, \quad
% \int_{p} \equiv \frac{1}{V} \sum_{\{n_\mu\}}, \quad
 \int_{p'} \equiv \frac{1}{V} \sum_{\{n_\mu\}} 
 \Bigl(1 - \delta^{(4)}_{n,0} \Bigr).
\ee

We will compute at next-to-leading order in the $\epsilon$-expansion, 
including corrections of relative order ${\cal O}(\epsilon^2)$. 
It turns out that the physical pion decay constant and mass, 
$F_\pi, M_\pi$, differ from their leading order values, 
$F_{},2 m \Sigma/F^2$ (for $M = \mathop{\mbox{diag}}(m,m,m)$), only by terms 
of relative order ${\cal O}(\epsilon^4)$~\cite{hl2}, an effect which may thus
be ignored. This is a result of the fact that 
no higher order operators in the action 
(i.e.,~none of the $L_i$'s of Gasser and Leutwyler)
contribute at ${\cal O}(\epsilon^2)$. 
Similarly,  there are no higher order operators 
of ${\cal O}(\epsilon^2)$ in the chiral 
representation of $[O_{w}]_{\a\b\c\d}$ in \eq\nr{O_XPT}. 
This is also in contrast with the usual 
chiral expansion in infinite volume, where a plethora of new operators 
contribute at next-to-leading order~\cite{p4}. The fact that the
contamination from higher order operators is small,  
is simply a consequence of being
closer to the chiral limit. 

The propagator for the non-zero momentum modes $\xi$ follows
by expanding the parametrisation in~\eq\nr{U_0} in $\epsilon$, 
and inserting into \eq\nr{LE}: 
\be
 \Bigl\langle \xi_{\c\a}(x) \, \xi_{\d\b}(y) \Bigr\rangle  = 
 \fr12 \Bigl[\delta_{\c\b} \delta_{\d\a} G(x-y) - 
 \delta_{\c\a} \delta_{\d\b} E(x-y) \Bigr]\;, \la{gen_prop}
\ee
where
\be
 G(x) = \int_{p'} \frac{e^{i p \cdot x}}{p^2}. 
\ee
In the chiral case, $E(x) = G(x)/N_f$, but 
we keep everywhere $E(x)$ completely general. The reason is that then the 
form of~\eq\nr{gen_prop} is general enough to contain also the propagator of 
the replica formulation of quenched chiral perturbation theory~\cite{ds,ddhj}, 
and thus we can already include 
the main ingredients needed in~\se\ref{se:quench}.

Where encountered, 
ultraviolet divergences are treated with dimensional regularization. 
For later reference let us define~\cite{hl,h}, in particular, 
the two integrals appearing in the computation (see also~\cite{mlv}): 
\ba 
 && \int_{p'} \frac{1}{p^2} 
% = \frac{L_0}{L_1L_2L_3} \biggl[ 
% \frac{1}{12} + \sum_{\vec{p}'} \frac{1}{|\vec{p}|L_0}
% \biggl( 
% \frac{1}{\exp(|\vec{p}|L_0)-1} + \fr12
% \biggr)
% \biggr]_\rmi{$\msbar$} 
 \equiv -\frac{\beta_1}{V^{1/2}}, \la{b1} \\
% & & 
% \int_x \Bigl[\partial_0 \bar G(x) \partial_0 \bar G(x) - 
% \bar G(x) \partial_0^2 \bar G(x) - \bar G(0) \delta (x) \Bigr] \nn
 && \int_{p'} \biggl(\frac{2 p_0^2}{(p^2)^2} - \frac{1}{p^2} \biggr)  
 = L_0 \frac{{\rm d}}{{\rm d} L_0} \int_{p'} \frac{1}{p^2} 
% = \frac{L_0}{L_1L_2L_3} \biggl[ 
% \frac{1}{12} - \sum_{\vec{p}'}\frac{1}{4 \sinh^2 (|\vec{p}| L_0/2)}
% \biggr] 
 \equiv \frac{L_0}{L_1L_2L_3} k_{00}.  \la{k00}
\ea
Here $\beta_1, k_{00}$ are finite dimensionless numerical coefficients 
depending on the geometry of the box. Some values
for them are listed in Table~1.

%%%%%%%%%%%%%%%%%%%%%%%%%%%%%%%%%%% TABLE %%%%%%%%%%%%%%%%%%%%%%%%%%%%%%%%
\begin{table}[t]

\centering
\begin{tabular}{lll}
\hline
$T/L$ & $\beta_1$ & $k_{00}$ \\ \hline
32/32 & 0.14046   & $0.07023 = \beta_1/2$ \\ 
32/28 & 0.13872   & 0.07826 \\ 
32/24 & 0.13215   & 0.08186 \\ 
32/20 & 0.11689   & 0.08307 \\ 
32/16 & 0.08360   & 0.08331 \\ \hline
\end{tabular}

\caption[a]{\it Some numerical values for $\beta_1$, $k_{00}$, 
defined in~\eqs\nr{b1}, \nr{k00}, for geometries of the type 
$L_0 \equiv T, L_1=L_2=L_3 \equiv L$.}

\end{table}
%%%%%%%%%%%%%%%%%%%%%%%%%%%%%%%%%%%%%%%%%%%%%%%%%%%%%%%%%%%%%%%%%%%%%%%%%%

%%%%%%%%%%%%%%%%%%%%%%%%%%%% SECTION %%%%%%%%%%%%%%%%%%%%%%%%%%%%%%%%%%%%%%%%
%
\section{Results}
\la{se:results}

\subsection{Charge -- charge expectation value}
\label{cc}

%%%%%%%%%%%%%%%%%%%%%%%%%%%% FIGURE %%%%%%%%%%%%%%%%%%%%%%%%%%%%%%%%%%%%%%%
\begin{figure}[t]
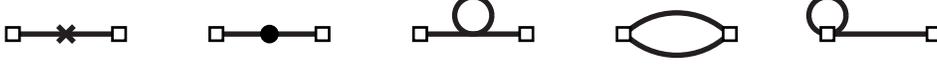


\begin{eqnarray*}
& &
\CTopomeas(\TLsc,\TLsc) \quad
\CTopomass(\TLsc,\TLsc) \quad 
%% \\
%% & & 
\CTopoin(\TLsc,\TLsc,\TAsc) \quad
\CTopocucu(\TAsc,\TAsc) \quad
\CTopocu(\TAsc,\TLsc) 
\end{eqnarray*}

\caption[a]{\it The ${\cal O}(\epsilon^2)$ graphs computed in \se\ref{cc}. 
An open box denotes a current (\eq\nr{cJmua}), a cross a ``measure 
term''~(cf.~ref.~\cite{gl}), and a filled circle a mass insertion.}
\la{fig:current}
\end{figure}
%%%%%%%%%%%%%%%%%%%%%%%%%%%%%%%%%%%%%%%%%%%%%%%%%%%%%%%%%%%%%%%%%%%%%%%%%%

We now proceed to apply the rules of the $\epsilon$-expansion to
the correlator ${\cal C}^{ab}(x_0)$, defined in~\eq\nr{Cab}. 
The result can in fact also be inferred from~\cite{h}, by summing together 
the expressions for the axial and vector flavour currents. 

The graphs contributing to ${\cal C}^{ab}(x_0)$ 
are shown in~\fig\ref{fig:current}. 
Apart from the graph including the mass insertion, 
current conservation guarantees that the result is independent of $x_0$. 
Indeed, we obtain
\ba
 {\cal C}^{ab}(x_0) & = & \Bigl(- \tr T^a T^b\Bigr) \frac{F_{}^2}{2 L_0} \nn
 & \times & \biggl\{ 1 + \frac{N_f}{F_{}^2}
 \biggl[ 
 \frac{\beta_1}{V^{1/2}} - \frac{L_0^2 k_{00}}{V} 
 \biggr]
 +\frac{2 \Sigma L_0^2}{N_{\! f} F_{}^2}\; 
 {\Bigl\langle \re \tr[M U_0 e^{i \theta/N_f}]\Bigr\rangle}_{\theta,U_0}\;
 h_1\Bigl( \frac{x_0}{L_0}\Bigr) 
 \biggr\},   \hspace*{1cm} \la{res_Cab}
\ea
where $\beta_1, k_{00}$ are from \eqs\nr{b1}, \nr{k00};
and~\cite{hl}
\be
 h_1(\tau) = \fr12 \Bigl[
 \Bigl(|\tau| - \fr12\Bigr)^2 - \fr{1}{12} \Bigr]. 
\ee
Finally, $\langle...\rangle_{\theta,U_0}$ denotes an average over
the zero-mode Goldstone manifold, 
\be
 \langle ... \rangle_{\theta,U_0} \equiv
 \frac{\int_{U_0} (...) \exp(V \Sigma \re \tr[M U_0 e^{i \theta/N_f}])}
 {\int_{U_0}\exp(V \Sigma \re \tr[M U_0 e^{i \theta/N_f}])},  
\ee
where $\int_{U_0}$ is an integration 
over SU($N_f$) according to the Haar measure.

%%%%%%%%%%%%%%%%%%%%%%%%%%%% FIGURE %%%%%%%%%%%%%%%%%%%%%%%%%%%%%%%%%%%%%%%
\begin{figure}[t]

\begin{center}
\epsfig{file=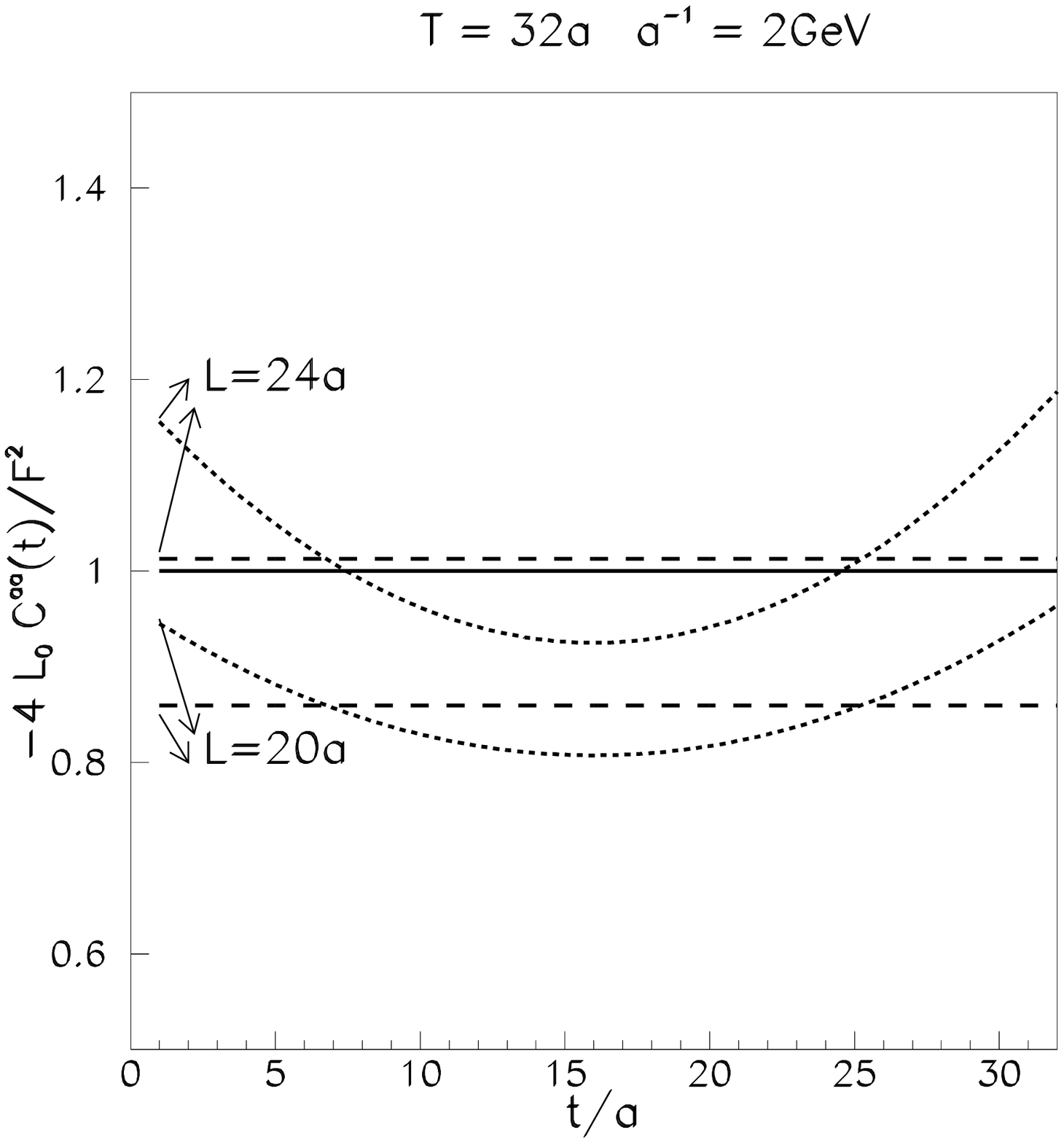,width=7cm}% 
%\hspace*{1cm}
\epsfig{file=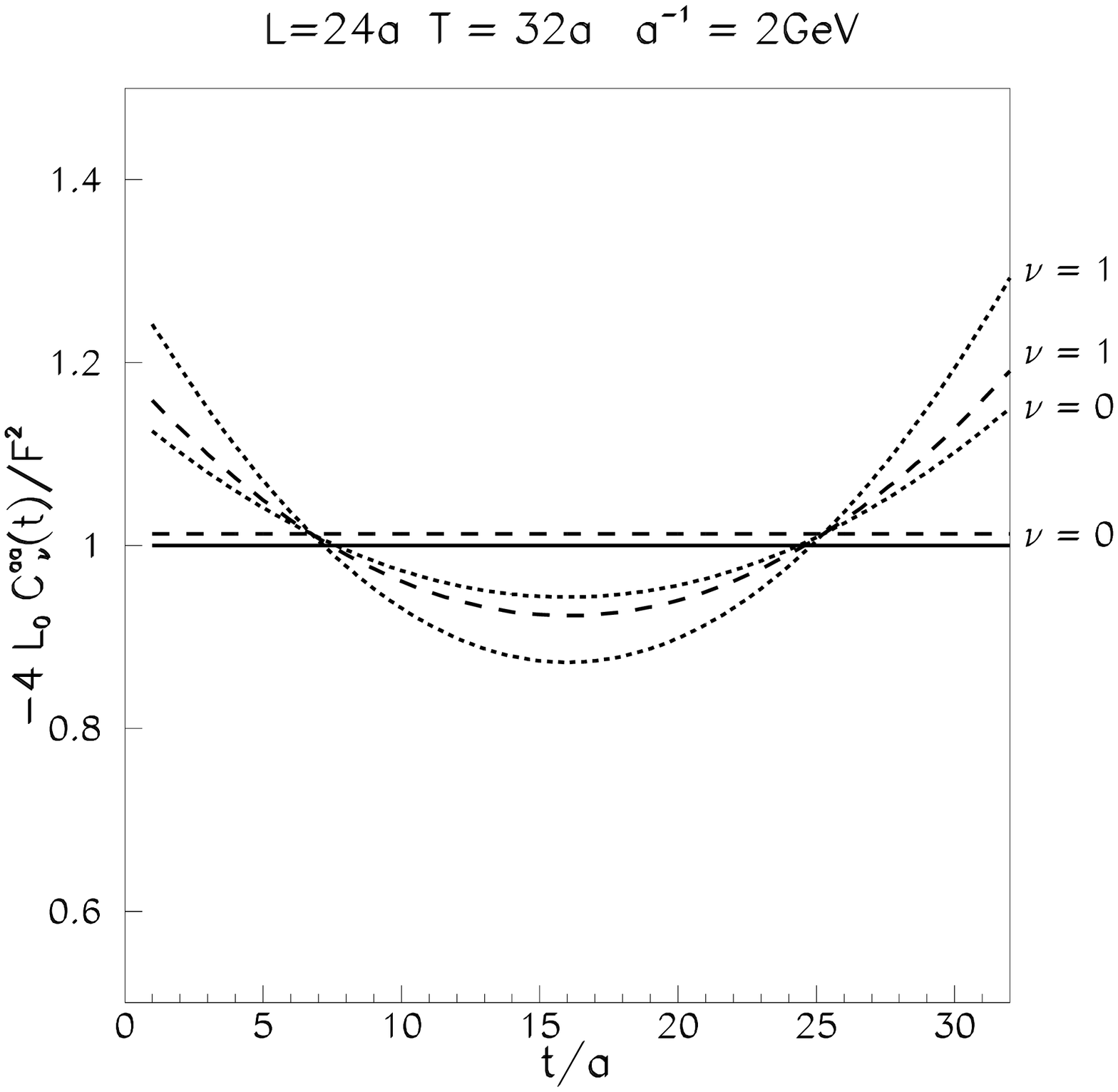,width=7cm}
\end{center}

\caption[a]{\it Left: 
The expression inside the curly brackets in~\eq\nr{res_Cab}, 
for $L_0=T$, $V = T L^3$, $a^{-1} = 2$ GeV, 
and $m=0$ MeV (dashed), $m=5$ MeV (dotted). 
We have assumed $\theta=0$, $F = 93$ MeV, $\Sigma = (250 \mbox{ MeV})^3$. 
The solid line is the tree-level result.
Right: The same observable, in an ensemble with a fixed
topological charge $\nu = 0,1$ (cf.~\se\ref{se:top}),~for~$L=24a$.}

\la{fig:current_numerics}
\end{figure}
%%%%%%%%%%%%%%%%%%%%%%%%%%%%%%%%%%%%%%%%%%%%%%%%%%%%%%%%%%%%%%%%%%%%%%%%%%%

As a simple explicit example,
let us assume a box with $L_0 = L_1 = L_2 = L_3$, a mass matrix 
$M  = \mathop{\mbox{diag}}(m,m,m)$, and a phase $\theta=0$. 
Taking furthermore into account that at the present
order ${\cal O}(\epsilon^2)$, $F_\pi = F_{}$, 
$M_\pi^2 = 2 m \Sigma/F_{}^2$,
the full result may be expressed as
\ba
 &&  {\cal C}^{ab}(x_0)
 %%\nn 
 %%&& 
 = \Bigl(- \tr T^a T^b\Bigr)
 \frac{F_\pi^2}{2 L_0} \biggl\{ 
 1 + \frac{1}{F_\pi^2 L_0^2} \biggl[\fr32 \beta_1
% \nn
% & + & 
 + u \; \Sigma_{\theta=0}(u/{2}) \, 
 h_1\Bigl(\frac{x_0}{L_0}\Bigr) \biggr]
 \biggr\}, \hspace*{1cm} \la{exp_res_0}
\ea
where~\cite{gl,h,hl2}
\ba
 u & = & M_\pi^2 F_\pi^2 L_0^4, \la{def_u} \\
 \Sigma_\theta({u}/{2}) & = &  
 \frac{1}{N_f} \Bigl\langle
 \re \tr U_0 e^{i \theta/N_f}
 \Bigr\rangle_{\theta,U_0} =
 \frac{2}{N_f}\frac{\partial}{\partial u}
 \ln \int_{U_0} e^{(u/2) \re \tr U_0 \exp({i \theta/N_f})} \;,  \\
 \Sigma_{\theta=0}(u/2)& \approx &  
 \left\{
 \begin{array}{cc}
 u/(4 N_f), & u \ll 1 \\
 1, & u \gg 1 
 \end{array}
 \right. \!\!\!\! . \hspace*{0.5cm} \la{def_Cu}
\ea
% The accuracy of the expression $u\, C(u/2)$ could also be 
% improved to the next order in the $\epsilon$-expansion, simply by 
% modifying the expression of $u$ in terms of $M_\pi,F_\pi$ with a known 
% correction of relative order ${\cal O}(\epsilon^2)$~\cite{hl2}.
Note that in the notation of~\cite{hl2}, 
$u\, \Sigma_{\theta=0}(u/2) = u^2 I_1(u)/(4 N_f)$.

For $N_f=3$, a numerical determination of~$I_1(u)$ 
has been given in~\cite{hl2}. Using this result, 
we show in~\fig\ref{fig:current_numerics} examples for 
two asymmetric lattices, 
$32\times 24^3, 32\times 20^3$, and a lattice spacing 
$a^{-1} = 2$ GeV. The function in~\eq\nr{res_Cab} has been 
normalised to its (constant) tree-level value.  
The next-to-leading order correction is observed 
to become dangerously large for $L/a \lsim 20$
(i.e., $L \lsim 2$ fm). 

\subsection{Charge -- charge -- weak operator ${\cal O}_{w}$}
\label{cco}

We then move to 
${[{\cal C}_{w}]}^{ab}_{\a\b\c\d}(x_0,y_0)$, defined in~\eq\nr{Cab_abcd}.
To this end we evaluate the graphs in~\fig\ref{fig:operator}. 
Let us mention that we are ignoring
disconnected diagrams, since they only lead to trace parts, 
$\sim \delta^{ab}\delta_{\c\a}\delta_{\d\b}$, 
$\delta^{ab}\delta_{\c\b} \delta_{\d\a}$, 
which would vanish in 
any case after the projection to $[\hat {\cal O}_{w}]^+_{\a\b\c\d}$.
The non-trivial flavour structures 
arising from~\fig\ref{fig:operator}
always appear in one of the two combinations, 
\ba
 [ \Delta^{(1)} ]^{ab}_{\a\b\c\d} & \equiv & 
 T^a_{\c\a} T^b_{\d\b} + T^a_{\d\b} T^b_{\c\a} \equiv
 T^{\{a}_{\c\a} T^{b\}}_{\d\b}, \la{delta1} 
 \\
 {[ \Delta^{(2)} ]}^{\raise-0.6ex\hbox{$\scriptstyle ab$}}_{\a\b\c\d} 
 & \equiv & 
 [ \Delta^{(1)} ]^{ab}_{\a\b\d\c} - \fr12 
 \Bigl( 
 \delta_{\c\b} \{ T^a,T^b \}_{\d\a} +
 \delta_{\d\a} \{ T^a,T^b \}_{\c\b} 
 \Bigr).  \la{delta2}
\ea
After lengthy but straightforward algebra, we obtain\footnote{%
 The full expression before the volume average 
 $\int_{\vec{x}} (...)$ can also be obtained from the authors on request.}
\ba
 {[{\cal C}_{w}]}^{ab}_{\a\b\c\d}(x_0,y_0) 
 \!\! & = & \!\!  -  \frac{F_{}^4}{4 L_0^2}
 \biggl\{ 
 \Delta^{(1)}
 +\Bigl(N_f \Delta^{(1)} + \Delta^{(2)}\Bigr)
 \frac{2}{F_{}^2}
 \biggl[ 
 \frac{\beta_1}{V^{1/2}} - \frac{L_0^2 k_{00}}{V} 
 \biggr] \nn
 & & \hspace*{0.8cm} +  
 \Delta^{(1)} 
 \frac{2 \Sigma L_0^2}{N_{\! f} F_{}^2} \; 
 {\Bigl\langle \re \tr[M U_0 e^{i \theta/N_f}]\Bigr\rangle}_{\theta,U_0}\;
 \biggl[ 
 h_1\Bigl(\frac{x_0}{L_0}\Bigr)+ 
 h_1\Bigl(\frac{y_0}{L_0}\Bigr)
 \biggr]\biggr\}, \hspace*{1cm} \la{Cabcd}
\ea
where we have for clarity omitted the indices from 
$[\Delta^{(1)}]^{ab}_{\a\b\c\d}, [\Delta^{(2)}]^{ab}_{\a\b\c\d}$.
Apart from the graph including the mass 
insertion, current conservation guarantees again 
that the result is independent of $x_0,y_0$.

%%%%%%%%%%%%%%%%%%%%%%%%%%%% FIGURE %%%%%%%%%%%%%%%%%%%%%%%%%%%%%%%%%%%%%%%
\begin{figure}[t]
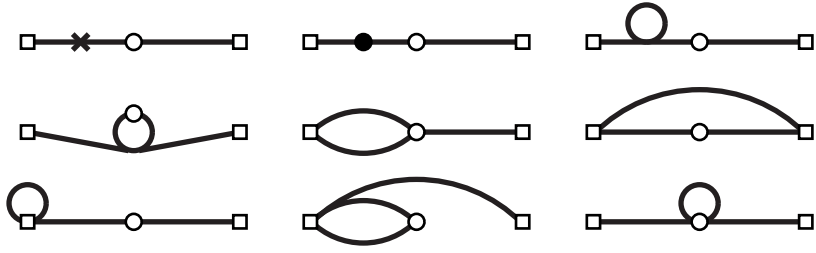


\begin{eqnarray*}
& &
\Topomeas(\TLsc,\TLsc,\TLsc) \quad
\Topomass(\TLsc,\TLsc,\TLsc) \quad
\Topoin(\TLsc,\TLsc,\TLsc,\TAsc) \\
& & 
\Topoinop(\TLsc,\TLsc,\TAsc) \quad
\Topocuop(\TAsc,\TAsc,\TLsc) \quad
\Topocucu(\TAsc,\TLsc,\TLsc) \\
& & 
\Topocu(\TAsc,\TLsc,\TLsc) \quad
\Topocucuop(\TAsc,\TAsc,\TAsc) \quad
\Topoop(\TAsc,\TLsc,\TLsc) 
\end{eqnarray*}

\caption[a]{\it The ${\cal O}(\epsilon^2)$ graphs computed in \se\ref{cco}. 
An open circle denotes the weak operator in \eq\nr{O_XPT}, otherwise 
the notation is as in~\fig\ref{fig:current}.}
\la{fig:operator}
\end{figure}
%%%%%%%%%%%%%%%%%%%%%%%%%%%%%%%%%%%%%%%%%%%%%%%%%%%%%%%%%%%%%%%%%%%%%%%%%%

Once the flavour structure is projected onto the 27-plet according 
to~\ref{app:su3}, we get our final result,  
\be
 [{\cal C}_{27}]^{ab}_{\a\b\c\d}(x_0,y_0) \equiv  
 \int_{\vec x} \int_{\vec y} \langle {\cal J}^a_0(x) 
 [\hat {\cal O}_{w}]_{\a\b\c\d}^+(0) {\cal J}^b_0(y) \rangle \;.
 \la{calc} 
\ee
It is directly obtained from \eq\nr{Cabcd}, 
by simply making the replacements
\ba
 [\Delta^{(i)}]^{ab}_{\a\b\c\d} \! & \to & \! 
 \hat \Delta^{ab}_{\a\b\c\d} \;, \quad i=1,2\;,
\ea
where
\ba
 \hat \Delta^{ab}_{\a\b\c\d} \! & \equiv & \! \fr12
 \Bigl( 
 T^{\{a}_{\c\b} T^{b\}}_{\d\a} +
 T^{\{a}_{\c\a} T^{b\}}_{\d\b}  
 \Bigr)  + \frac{1}{20}  
 \Bigl( \delta_{\c\b}\delta_{\d\a} + 
 \delta_{\c\a} \delta_{\d\b} \Bigr) \tr T^a T^b
 \la{full_structure} \\
 \! & - & \! \frac{1}{10} \Bigl( 
 \delta_{\c\b} \{ T^a,T^b \}_{\d\a} + 
 \delta_{\d\a} \{ T^a,T^b \}_{\c\b} + 
 \delta_{\c\a} \{ T^a,T^b \}_{\d\b} + 
 \delta_{\d\b} \{ T^a,T^b \}_{\c\a} \Bigr). \nonumber 
\ea

As a simple explicit example,
we again assume a box with $L_0 = L_1 = L_2 = L_3$, a mass matrix 
$M  = \mathop{\mbox{diag}}(m,m,m)$, a phase $\theta=0$, $N_f=3$, 
and choose generators such that  
\be
 T^a_{ij} \equiv \delta_{i\2}\delta_{j\3}, \quad
 T^b_{ij} \equiv \delta_{i\1}\delta_{j\2}. \la{indices}
\ee
We also choose the physical indices for $[\hat {\cal O}_{w}]_{\a\b\c\d}^+$, 
according to~\eq\nr{formofO}. Then $\hat\Delta^{ab}_{\3\2\2\1} = 2/5$, and 
the full result is
\be
 {[ {\cal C}_{27}]}^{ab}_{\3\2\2\1}(x_0,y_0)
 %%\nn 
 %%&& 
 = - \frac{F_\pi^4}{4 L_0^2}  \fr25 \biggl\{ 
 1 + \frac{1}{F_\pi^2 L_0^2} \biggl[ 4 {\beta_1}
% \nn
% & + & 
 + u \, \Sigma_{\theta=0}(u/{2}) \,
 \Bigl[ 
 h_1\Bigl(\frac{x_0}{L_0}\Bigr)+ 
 h_1\Bigl(\frac{y_0}{L_0}\Bigr) \Bigr]
 \biggr] \biggr\}, \hspace*{0.9cm} \la{exp_res}
\ee
where the notation is as in~\eq\nr{exp_res_0}.

In~\fig\ref{fig:osigma_numerics} we show the predictions of~\eq\nr{Cabcd} 
for this index choice, normalised to the tree-level value, for two 
asymmetric volumes. 
Since there is only one 27-plet operator in~\eq\nr{Lw_QCD_su3}, 
a measurement of $g_{27}$ can then 
be obtained through the matching of the 
chiral three-point function with the corresponding lattice QCD measurement:
\be
 \fr53 g_{27} {[{\cal C}_{27}]^{ab}_{\3\2\2\1}(x_0,y_0)} = h_{w}^+  
 \int_{\vec x} \int_{\vec y} \langle {J}^a_0(x) 
 [\hat {O}_{w}]_{\3\2\2\1}^+(0) {J}^b_0(y) \rangle \;. 
 \; 
\ee

%%%%%%%%%%%%%%%%%%%%%%%%%%%% FIGURE %%%%%%%%%%%%%%%%%%%%%%%%%%%%%%%%%%%%%%%
\begin{figure}[t]

\begin{center}
\epsfig{file=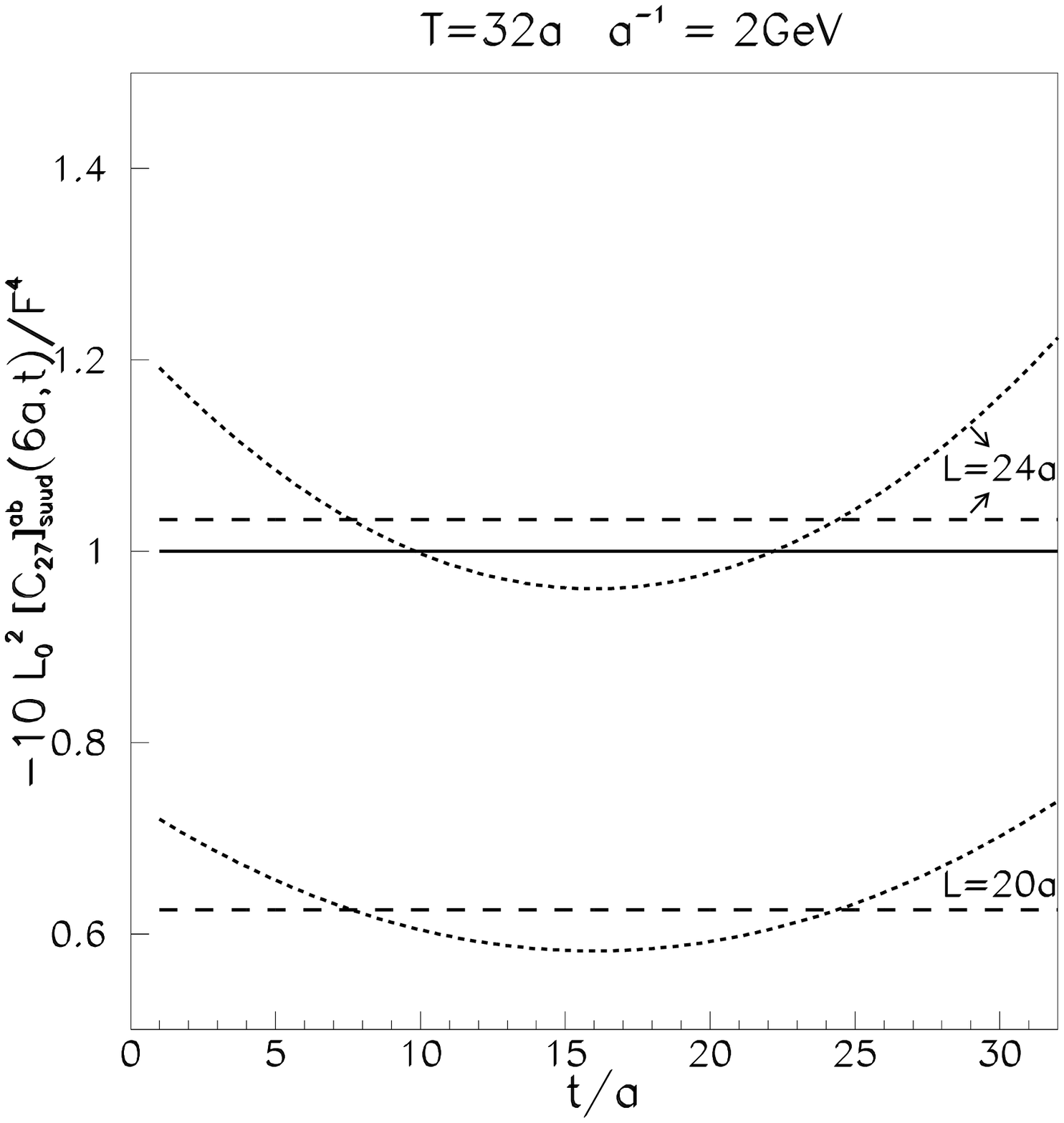,width=7cm}% 
\epsfig{file=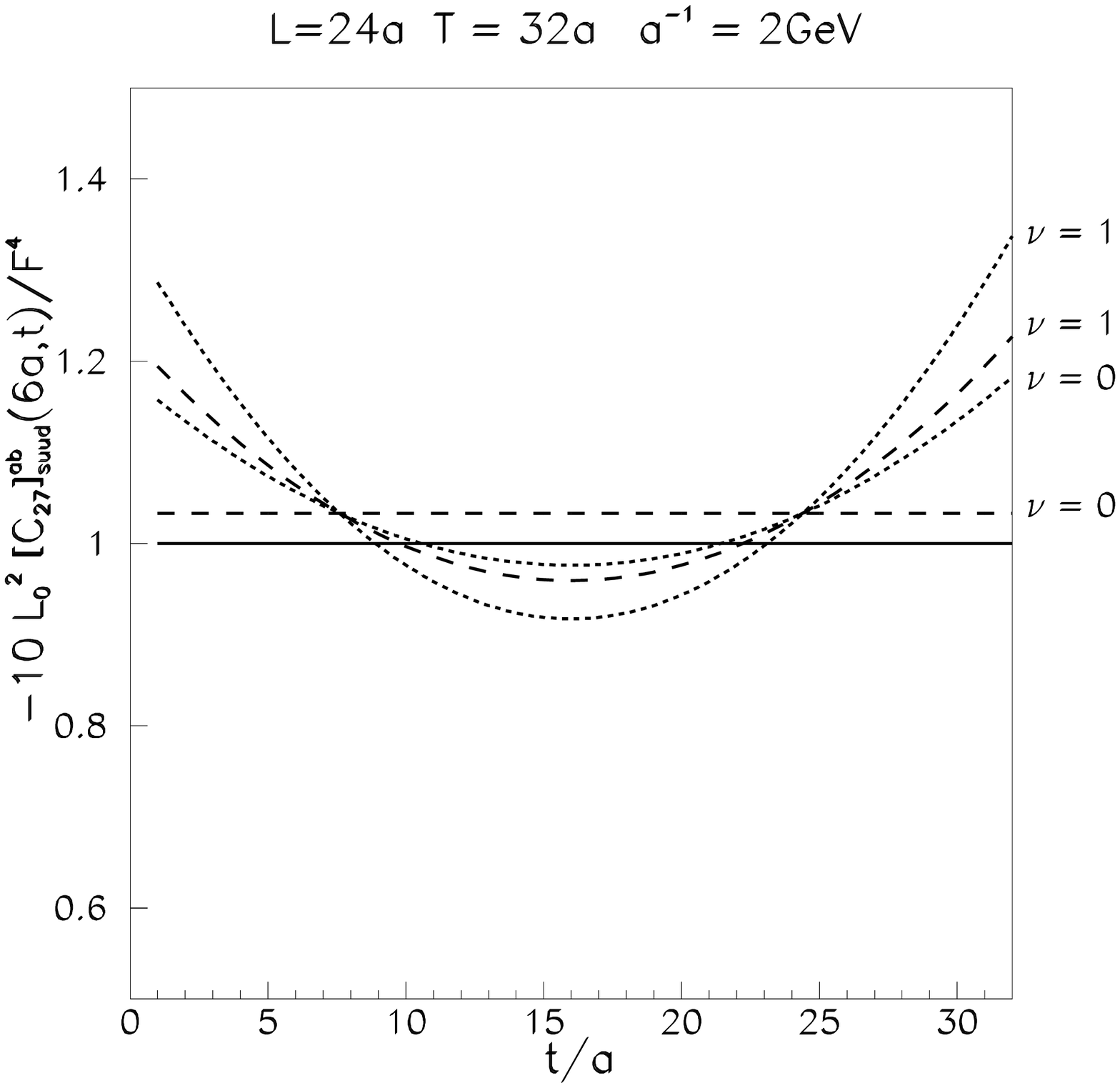,width=7cm}
\end{center}

\caption[a]{\it Left: 
The result of~\eq\nr{Cabcd} for the index
choice in~\eqs\nr{indices}, \nr{exp_res}, normalised to the tree-level value
(solid line), 
for the same parameters as in~\fig\ref{fig:current_numerics}.
The upper set is for $L=24a$, the lower for $L=20a$.
The dashed and dotted
lines correspond to $m=0$ MeV and $m=5$ MeV, respectively.
Right: The same observable, in an ensemble with a fixed
topological charge $\nu = 0,1$ (cf.~\se\ref{se:top}),~for~$L=24a$.}

\la{fig:osigma_numerics}
\end{figure}
%%%%%%%%%%%%%%%%%%%%%%%%%%%%%%%%%%%%%%%%%%%%%%%%%%%%%%%%%%%%%%%%%%%%%%%%%%%

\subsection{Fixed topology}
\la{se:top}

The results above were obtained in a fixed $\theta$-vacuum. We can also  
perform a Fourier transform in $\theta$ to obtain 
averages in sectors of ``fixed topology'' $\nu$~\cite{ls}. This is interesting 
because in the quenched theory we expect to find poles in $m$ in fermion 
propagators, which become dominant in the $\epsilon$-regime
when $m \Sigma V \ll 1$.  
In the $\theta$-vacuum there are no such 
poles in the full theory, because topological configurations 
are strongly suppressed by the fermion determinant. However, when 
considering averages in sectors of non-zero fixed topology, the same poles
are expected to appear in the quenched and the full theories. 
It is quite remarkable that these poles appear also in the 
corresponding effective chiral theories! 
Even though their presence does not 
affect the counting rules of the $\epsilon$-expansion, 
because $m \Sigma V$ is formally counted
as a quantity of ${\cal O}(1)$, they obviously modify the chiral limit.
The question we want to address here is whether there are such poles 
in the observables of \eqs\nr{Cab}, \nr{Cab_abcd}. 

An observable in the sector of topological charge $\nu$, $f_\nu$, 
can be obtained 
from the observable in a $\theta$-vacuum, $f_\theta$, by 
\be
 f_\nu = \frac{1}{2\pi} \int_0^{2\pi} {\rm d} 
 \theta e^{-i \nu \theta} f_\theta.
\ee
In particular, assuming again 
$M  = \mathop{\mbox{diag}}(m,m,m)$ and defining
\be
 Z_\theta(u/2) = \int_{U_0} e^{(u/2) \re \tr [U_0 \exp(i \theta/N_f)]}, \quad
 Z_\nu(u/2) = \frac{1}{2\pi} \int_0^{2\pi} {\rm d} \theta e^{-i \nu \theta}
 Z_\theta(u/2), 
\ee
the combination appearing 
in~\eqs\nr{res_Cab}, \nr{Cabcd} gets replaced as 
\ba  
 %% \frac{1}{2\pi} \int_0^{2\pi} {\rm d} \theta e^{-i \nu \theta} 
 %% \frac{1}{N_f} \Bigl\langle  \re \tr [U_0 e^{i \theta/N_f}]
 %% \Bigr\rangle_{\theta,U_0}
 \Sigma_\theta(u/2) =  
 \frac{2}{N_f} 
 \frac{\partial}{\partial u} \ln Z_\theta(u/2)
 & \longrightarrow & \frac{2}{N_f} 
 \frac{\partial}{\partial u} \ln Z_\nu(u/2) \equiv \Sigma_\nu(u/2). 
 \hspace*{0.8cm}
\ea
The function $Z_\nu$ is known~\cite{brower,ls} to be
\be
 Z_\nu(u/2) = \det[I_{\nu+j-i}(u/2)],  
\ee 
where the determinant is taken over an $N_f \times N_f$ matrix, whose 
matrix element $(i,j)$ is the modified Bessel function $I_{\nu+j-i}$. 

Thus, at fixed topology the results 
corresponding to \eqs\nr{exp_res_0}, \nr{exp_res} 
are obtained by the substitution 
$\Sigma_\theta(u/2) \rightarrow \Sigma_\nu(u/2)$. For small 
and large $u$ we have (independent of $N_f$),
\be
 \Sigma_\nu(u/2) \approx 
 \left\{
 \begin{array}{cc}
  2 {|\nu|}/{u}, & u \ll 1 \\
  1, & u \gg 1 
 \end{array}
 \right. \!\!\!\! . \la{Cnu}
\ee
As expected the low mass behaviour ($u\ll 1$) is drastically modified 
with respect to that in~\eq\nr{def_Cu}. This implies that  
even though the correlators remain finite for $m\to 0$
(i.e., there are no poles, because $\Sigma_\nu(u/2)$ is 
multiplied by $u$), their time
dependence does not vanish. This is illustrated for 
${\cal C}^{ab}(x_0)$ in~\fig\ref{fig:current_numerics}
and for $[{\cal C}_{27}]^{ab}_{\3\2\2\1}(x_0,y_0)$ 
in~\fig\ref{fig:osigma_numerics}.

%%%%%%%%%%%%%%%%%%%%%%%%%%%%%%%%%%%%%%%%%%%%%%%%%%%%%%%%%%%%%%%%%%%%%%%%
\subsection{Normalised correlators}
\la{se:normalised}

%%%%%%%%%%%%%%%%%%%%%%%%%%%% FIGURE %%%%%%%%%%%%%%%%%%%%%%%%%%%%%%%%%%%%%%%
\begin{figure}[t]

\begin{center}
\epsfig{file=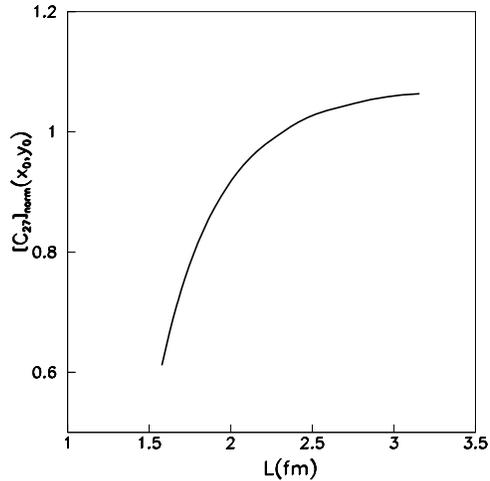,width=7cm}
\end{center}

\caption[a]{\it 
The result of~\eq\nr{norm_Cabcd}, 
for the same parameters as in~\fig\ref{fig:current_numerics}.
At this order, the outcome is independent of $x_0,y_0$.}

\la{fig:norm_numerics}
\end{figure}
%%%%%%%%%%%%%%%%%%%%%%%%%%%%%%%%%%%%%%%%%%%%%%%%%%%%%%%%%%%%%%%%%%%%%%%%%%%

The predictions of the previous sections depend 
at leading order on the chiral theory parameter
$F_{}$, show rather bad convergence at small volumes, $L\lsim 2$~fm, 
and have (for a non-vanishing mass, 
as well as for a non-zero topological charge) 
a non-trivial dependence on $x_0,y_0$. All these dependences
are a nuisance for the determination of $g_{27}$. 
Fortunately, there seems to be a large cancellation
if we normalise the three-point function
$[{\cal C}_{27}]^{ab}_{\3\2\2\1}$ by 
two charge -- charge correlators, at least at 
the present order in the $\epsilon$-expansion. 

More precisely, let us again choose the indices 
in~\eqs\nr{indices}, \nr{exp_res}, and denote by ${\cal C}^{a a^\dagger}(x_0)$ 
a charge -- charge correlator obtained by using the 
generators $T^a, (T^a)^\dagger$ in the currents. Expanding the 
denominators, we then obtain that to relative order ${\cal O}(\epsilon^2)$, 
\be
 [ {\cal C}_{27} ]_\rmi{norm} (x_0,y_0) \equiv 
 - \fr52 \frac{[{\cal C}_{27}]^{ab}_{\3\2\2\1}(x_0,y_0)}
 {{\cal C}^{aa^\dagger}(x_0) {\cal C}^{bb^\dagger}(y_0)} = 
 1 +  \frac{2}{F_{}^2} \biggl[ 
 \frac{\beta_1}{V^{1/2}} - \frac{L_0^2 k_{00}}{V} 
 \biggr]. 
 \la{norm_Cabcd}
\ee
The same result holds at fixed topology, as defined in~\se\ref{se:top}.

Thus, the time, quark mass, and topology dependences cancel
completely in the ratio of~\eq\nr{norm_Cabcd}! The next-to-leading
order correction is also numerically smaller than the
corresponding corrections in the numerator and denominator separately. 
This result is illustrated in~\fig\ref{fig:norm_numerics} as a function 
of the spatial volume, for a fixed time-like extent $T = 3.2$~fm.

To summarise, the optimal method for determining $g_{27}$ 
would appear to be from the equality
\be
 g_{27}  = \fr35 h_{w}^+ { {[{C}_{27}]_\rmi{norm} (x_0,y_0)} \over 
 {[{\cal C}_{27} ]_\rmi{norm} (x_0,y_0)}}
 \;, \la{match_final}
\ee
where $[ {C}_{27} ]_\rmi{norm}$ is the QCD-correspondent 
for the expectation value in~\eq\nr{norm_Cabcd}. The independence of 
the outcome on the volume, quark masses, $x_0,y_0$, and topological 
charge, serves as a test of whether the regime of applicability
of~\eq\nr{match_final} has been reached. 

%%%%%%%%%%%%%%%%%%%%%%%%%%%%% SECTION %%%%%%%%%%%%%%%%%%%%%%%%%%%%%%%%%%%%
\section{The quenched case}
\la{se:quench}

%%%%%%%%%%%%%%%%%%%%%%%%%%%%%%%%%%%%%%%%%%%%%%%%%%%%%%%%%%%%%%%%%%%%%%%%%%%%%
%
\subsection{The basic setup}

Due to the numerical cost of dynamical Ginsparg-Wilson fermions, 
practical lattice simulations will, for a while still, have to resort 
to the quenched approximation. It is therefore of interest to study
how the results of the previous sections are expected to be affected
by quenching. The tool for this is quenched chiral perturbation theory, 
applied to the $\epsilon$-regime. Previous results in this setup 
exist for the quark condensate~\cite{dotv}, and the scalar and 
pseudoscalar~\cite{ddhj} as well as flavoured vector 
and axial-vector~\cite{dhjll} two-point functions. 

There are two approaches to quenched QCD, believed to be equivalent: 
the so-called supersymmetric (SUSY) formulation~\cite{bg,sharpe} and 
the so-called replica method~\cite{ds,ddhj}. In the former, bosonic ``ghost'' 
quarks are introduced in order to cancel the effects of the physical 
quarks; in the latter, the computation is carried out by keeping
separately track of the $N_v$ ``valence'' quarks appearing in 
the external sources, and the $N_f$ dynamical quarks, and the 
quenched limit is obtained by taking $N_f \rightarrow 0$ for 
a fixed $N_v\neq 0$. 

The two methods are formally equivalent 
at the quark level, however their low-energy effective theories 
appear to be quite different. Assuming that the naive chiral 
symmetries of these models, U$(N_f)_L \times $U($N_f)_R$ in the replica 
case and the graded U($N_v|N_v)_L \times $U($N_v|N_v)_R$ 
in the SUSY formulation, are broken spontaneously by the 
formation of a quark condensate to the corresponding vector 
subgroups, the low-energy degrees of freedom are the resulting 
Goldstone bosons, whose dynamics can be described by 
chiral Lagrangians, at energies below the typical confinement scale.
The field variables of the chiral Lagrangians 
are matrices parametrising the Goldstone manifolds.

There is one important difference with respect to full QCD, 
though: the field associated with the singlet axial 
rotation ($\sim \eta'$) does not decouple from the low-energy dynamics. 
This is true both in the SUSY method~\cite{bg,sharpe}, as well as
in the replica: in order to have a sensible $N_f\rightarrow 0$ limit, 
the Goldstone manifold needs to be enlarged from 
SU($N_f$) to U($N_f)$. It is then easy 
to see that the decoupling of the singlet field and the limit 
$N_f\rightarrow 0$ do not commute~\cite{ds}. As a result, 
the chiral Lagrangian may contain all possible interactions 
involving the singlet, and to lowest order, the replica method has 
\ba
 {\cal L}_{\chi PT} \!\! & = & \!\! \frac{F^2}{4} \tr
 \Bigl[ \partial_\mu U \partial_\mu U^{-1} \Bigr] 
 - {\Sigma \over 2} \tr
 \! \Bigl[ U_{\theta} U M + M^\dagger U^{-1} U^{-1}_{\theta}\Bigr] %% \nn 
 %% & + & 
 + {m_0^2 \over 2 N_c} \Phi^2_0 + \frac{\alpha}{2N_{c}}
 (\partial_{\mu} \Phi_0)^2, \hspace*{0.5cm} %% \partial_{\mu}\Phi_0,
 \label{Lrep}
 \la{qLE}
\ea
where $\Phi_0 \equiv ({F_{}}/{{2}}) 
\tr [-i \ln (U)]$ and $U_\theta\equiv
\exp(i \theta {I_{N_v}} /N_v)$. Here, ${I_{N_v}}$ is the identity 
matrix in the valence subspace and zero elsewhere.
Obviously the couplings $F$ and $\Sigma$ 
need not be the same as in full QCD. 
In addition, new parameters related to axial singlet field, 
$m_0^2,\alpha$, have been introduced. 
In the SUSY formulation the first order chiral Lagrangian is 
the same, with the substitution $\tr \rightarrow \str$ and  
$U \in$~U($N_v|N_v)$ 
(or, more precisely, $U \in \widehat{\rm Gl}(N_v|N_v)$~\cite{z}). 

Since $\Phi_0$ is a singlet, there could in principle also be 
additional operators constructed with it in~\eq\nr{Lrep}. They have, 
however, been shown to be suppressed by additional powers of 
$1/N_c$~\cite{glc}, so we will neglect them in the following. 

Even though the low-energy Lagrangians of the replica and SUSY
theories have quite different dynamical degrees of freedom, 
it is believed that in perturbation theory all computations concerning
physical observables are equivalent~\cite{ds}. In 
the $\epsilon$-regime, however, a non-perturbative definition 
is needed for the zero-momentum integration. Traditionally this could
only be achieved with the SUSY formulation, but there have been 
recent developments whereby it is argued that replica integrations
can also be performed non-perturbatively, and agree with 
SUSY integrations~\cite{ksv}. Here we carry out the perturbative
part of the computation with the replica method, and return 
to the zero-mode integrations later on.

Provided that only sectors of fixed topology are considered, 
the rules of the $\epsilon$-expansion are as in~\eq(\ref{epsexp}). 
The massless non-zero mode Goldstone propagator is also needed. 
In the replica case, it is given by \eq(\ref{gen_prop}), with 
\ba
 E(x) & = & 
 \lim_{N_f \to 0}  
 \int_{p'} \frac{e^{i p \cdot x}}{p^4}
 \frac{(\alpha p^2 + m_0^2)/(2 N_c)}
%% \frac{\frac{\alpha p^2 + m_0^2}{2 N_c}}
 {1 + (N_f/p^2) (\alpha p^2 + m_0^2)/(2 N_c) } \nn 
%% {1 + \frac{N_f}{p^2}\frac{\alpha p^2 + m_0^2}{2 N_c} } \nn 
 & \equiv & \frac{\alpha}{2 N_c} G(x) + \frac{m_0^2}{2 N_c} F(x)\;,\quad
 F(x) \equiv \int_{p'} \frac{e^{i p \cdot x}}{p^4}. 
 \label{qprop}
\ea
However, as mentioned, our previous results were obtained 
with a completely general $E(x)$, and therefore we know that
\eqs\nr{res_Cab}, \nr{Cabcd} are independent of its form. 

%%%%%%%%%%%%%%%%%%%%%%%%%%%%%%%%%%%%%%%%%%%%%%%%%%%%%%%%%%%%%%%%%%%%%%%%%%%%%
%
\subsection{Currents and weak operators}

Let us then consider the left-handed currents, in the replica formulation. 
Since the current follows from the Lagrangian, cf.\ \eq\nr{cJmua}, 
and the additional degree of freedom  $\Phi_0$
is a flavour singlet, nothing changes with respect to the unquenched 
case at the present order:
\be
 %% \Bigl(J_\mu^a\Bigr)_\rmi{$\chi$PT} \equiv
 {\cal J}_\mu^{a,\rmi{quenched}} =  -i \frac{F^2}{2} T^a _{\a\b} \Bigl(
 \partial_\mu U U^{-1} \Bigr)_{\b\a},
 \la{qcJmua}
\ee
where the matrices $T^a$ are traceless and 
take non-zero values only in the valence sector. 

For the weak operators which do not directly follow from 
the Lagrangian, we have to be more careful. The general issue
is whether there are more operators once the larger 
symmetry group of the quenched theory is considered. Clearly, 
for instance, one could attach the singlet field $\Phi_0$ to 
any operator. As mentioned, we assume such terms to be 
suppressed by $1/N_c$ and ignore them. Another trivial issue 
is that trace parts of $(\partial_\mu U U^{-1})_{\a\c}$ vanish 
in the full theory but not in the quenched theory. However, there 
could in principle also be more drastic effects~\cite{gp}. 

To be systematic about the operators appearing, let us recall 
the symmetries that are relevant. The weak operators
we consider have indices
corresponding to left-handed valence flavours only. Therefore they
are singlets under the {\em full} right-handed symmetry
group\footnote{%
  This excludes the problematic operators
  considered in ref.~\cite{gp}, 
  $\sim [U \mathop{\mbox{diag}} (I_{N_v},-I_{N_v})\, U^{-1}]$,
  written here in the SUSY formulation.}, 
while they should have the correct symmetry properties under the 
left-handed {\em valence} subgroup. These requirements are sufficient to 
guarantee that the leading order 27-plet operator in the quenched theory 
is of the same type as in the unquenched case. 

Indeed, to get right-handed singlets under the full symmetry group, 
we are lead to the building blocks
\be
 \partial_\mu U U^{-1},  U \partial_\mu U^{-1} 
 \;\sim {\cal O}(\epsilon^2);\;\; 
 U M, \; M^\dagger U^{-1} \sim {\cal O}(\epsilon^4)\;,
 \la{ops_list}
\ee
which transform as fundamental $\otimes$ anti-fundamental 
under the valence subgroup SU($N_v)_L$, or SU$(N_v|N_v)_L$. 
We have also indicated the scalings of these operators
in the $\epsilon$-regime. The operators can be trivially decomposed 
into a sum of $\mathbf{3 \otimes 3^*}$, $\mathbf{1 \otimes 3^*}$, 
$\mathbf{3 \otimes 1}$ and $\mathbf{1 \otimes 1}$ irreducible
representations of the valence subgroup. To get a Lorentz invariant 
object with four flavour indices leads, as in the unquenched case, 
to a unique possibility up to and including ${\cal O}(\epsilon^6)$:
\be
 [{\cal O}_{w}]_{\a\b\c\d} = 
 \fr14 F^4 
 \Bigl(\partial_\mu U U^{-1}\Bigr)_{\c\a}
 \Bigl(\partial_\mu U U^{-1}\Bigr)_{\d\b}\;. 
 \la{O_qXPT}
\ee
The reduction to irreducible representations 
follows from~\ref{app:su3}.
Only the $\mathbf{3 \otimes 3^*}$ components of 
$\partial_\mu U U^{-1}$ contribute to the 27-plet, 
because the 27-plet cannot 
appear in the tensor product of less than four 
fundamentals/anti-fundamentals. In other words,
the operator $[{\hat {\cal O}}_{w}]^+_{\a\b\c\d}$ is symmetric in 
$\a\leftrightarrow\b$ and $\c\leftrightarrow\d$ 
and traceless in the SU($3)_L$ subgroup, and zero if any of the 
indices lies outside of this subgroup. 
The weak Hamiltonian reads then
\be
 {\cal H}_{w} = 
 2 \sqrt{2} G_F V_{ud} V^*_{us}
 \biggl\{
 \fr53 
 g^\rmi{quenched}_{27}
       [\hat {\cal O}_{w}]_{\3\2\2\1}^+ + ... \biggr\}+ \Hc \;,
 \la{Lw_qXPT}
\ee 
where we have indicated that the quenched coupling, 
$g^\rmi{quenched}_{27}$, does not need
to be the same as the analogous one in the unquenched theory.

In the case of the octets, on the other hand, 
the classification according to the valence 
group leads to ambiguities as discussed in~\cite{gp}. 

\vspace*{0.5cm}

Given that the currents
have the same form as in the full theory, 
the quenched two-point function ${\cal C}^{ab,\rmi{quenched}}$ 
can now be obtained by setting $N_f \to 0$ in the fixed-topology
version of~\eq\nr{res_Cab}, as this result 
is independent of the part $E(x)$ of the propagator. Alternatively,  
the result can be read from~\cite{dhjll}, by adding up 
the vector and axial-vector correlators and dividing by four:
\ba
 {\cal C}^{ab,\rmi{quenched}}(x_0) =  
 \Bigl(- \tr T^a T^b\Bigr) \frac{F^2}{2 L_0} 
 \times  \biggl\{ 1 + \frac{2 \Sigma L_0^2}{F^2} \frac{1}{N_v}\; 
 {\Bigl\langle \re \trp[M U_0 ]\Bigr\rangle}_{\nu,U_0}\;
 h_1\Bigl( \frac{x_0}{L_0}\Bigr) 
 \biggr\},   \hspace*{0.2cm} \la{res_Cqab}
\ea
where $\trp$ denotes the trace over the valence subgroup, with $N_v=3$.  
Let us stress that the absence of $E(x)$ guarantees that 
there is no dependence on the singlet couplings $m_0^2$ and $\alpha$, 
which appear in the non-zero mode propagator of \eq(\ref{qprop}). 
Another interesting point to note is that the terms proportional 
to $N_f$ of the unquenched result were largely responsible for the large 
corrections in realistic volumes, while they are absent now. This seems to 
imply that the $\epsilon$-expansion converges better in the quenched case.

Similarly, the result for the three-point function can 
be directly extracted from
the fixed-topology version of~\eq\nr{Cabcd}:
\ba
 [{\cal C}_{w}]^{ab,\rmi{quenched}}_{\a\b\c\d}(x_0,y_0) 
 \!\! & = & \!\! -  \frac{F^4}{4 L_0^2}
 \biggl\{ 
 \Delta^{(1)}
 +\Delta^{(2)}
 \frac{2}{F^2}
 \biggl[ 
 \frac{\beta_1}{V^{1/2}} - \frac{L_0^2 k_{00}}{V} 
 \biggr] \nn
 \!\! & &  \hspace*{0.2cm} +  
 \Delta^{(1)} 
 \frac{2 \Sigma L_0^2}{ F^2} \frac{1}{N_{\! v}} \; 
 {\Bigl\langle \re \trp[M U_0 ]\Bigr\rangle}_{\nu,U_0}\;
 \biggl[ 
 h_1\Bigl(\frac{x_0}{L_0}\Bigr)+ 
 h_1\Bigl(\frac{y_0}{L_0}\Bigr)
 \biggr]\biggr\}, \hspace*{1cm} \la{qCabcd}
\ea
where we have omitted the indices from 
$[\Delta^{(1)}]^{ab}_{\a\b\c\d}, [\Delta^{(2)}]^{ab}_{\a\b\c\d}$, 
given in \eqs(\ref{delta1}), (\ref{delta2}). Again, the 
unphysical axial singlet couplings $m_0^2,\alpha$ do not 
appear at this order. 

An important point to stress is that the zero-mode integrals
appearing in~\eqs\nr{res_Cqab}, \nr{qCabcd} are identical. 
Therefore, if we make the index choice of~\eqs\nr{indices}, \nr{exp_res}
and consider the ratio of \eq(\ref{norm_Cabcd}), the results 
are the same in the unquenched and quenched cases:
\ba
 [{\cal C}_{27}]^{\rmi{quenched}}_\rmi{norm} (x_0,y_0) 
 & \equiv &  
 - \fr52 \frac{[{\cal C}_{27}]^{ab,\rmi{quenched}}_{\3\2\2\1}(x_0,y_0)}
 {{\cal C}^{aa^\dagger,\rmi{quenched}}(x_0) 
  {\cal C}^{bb^\dagger,\rmi{quenched}}(y_0)} \nn
 & =  & 
 1 +  \frac{2}{F_{}^2} \biggl[ 
 \frac{\beta_1}{V^{1/2}} - \frac{L_0^2 k_{00}}{V} 
 \biggr] + {\cal O}(\epsilon^4) \;.
\ea 

%%%%%%%%%%%%%%%%%%%%%%%%%%%%%%%%%%%%%%%%%%%%%%%%%%%%%%%%%%%%%%%%%%%%%%%%%%%%%
%
\subsection{The quenched chiral condensate}

We end by discussing the actual value of the zero-mode integral 
in~\eqs\nr{res_Cqab}, \nr{qCabcd} for degenerate masses 
$M  =  \mathop{\mbox{diag}}(m,m,m)$: 
\be
 \Sigma_\nu^\rmi{quenched} \equiv 
 \frac{1}{2N_v} \left\langle \trp [ U_0 + U_0^{-1} ] \right\rangle_{\nu, U_0}  
%% \left\langle {\rm Re}[({U_0})_{11}] \right\rangle_{U_0,\nu} 
 \;. 
 \label{zero}
\ee
This is just the quark condensate, and the value is well 
known, for $N_v=1$~\cite{dotv}. It is usually assumed that 
the outcome should not depend on $N_v\geq 1$. What we wish 
to do here is to check this explicitly for $N_v=2,3$, by using 
the recent results from~\cite{ksv}:
\ba
 Z_{\nu,N_v}[J] \equiv \lim_{N_f\rightarrow 0} \int_\rmi{$U_0\in$ U($N_f$)} 
 {\det}^\nu U_0 e^{ \Sigma V \re 
 {\rm Tr}[M_J U_0]} 
 \sim \frac{\det\left[{\cal I}_{\nu,ij}(\mu_i)\right]_{i,j=1,...,2 N_v}}
 {\prod_{j>i=1}^{N_v} (\mu_j^2-\mu_i^2)},
 \label{sv}
\ea
where the $N_f \times N_f$ matrix $M_J$ is  
$M_J \equiv \mathop{\mbox{diag}}(m+J_1,...,m+J_{N_v},m,...)$, and 
$\mu_i = (m + J_i) \Sigma V$, $\mu = m \Sigma V$, 
together with 
\ba
 {\cal I}_{\nu,ij}(\mu_i) 
 & \equiv &  
 \left\{ \begin{array}{ll} 
 \mu_i^{j-1} I_{\nu+j-1}(\mu_i), 
 & i=1,...,N_v, \\  
 (-1)^{j-i+N_\nu}\mu^{j-1} 
 K_{\nu+j-i+N_\nu}(\mu), & i=N_v+1,...,2 N_v.
 \end{array} \right.
\ea
The derivative of the logarithm of this partition 
function with respect to $J_1$, evaluated at $J_i=0$, gives 
the required integral in~\eq(\ref{zero}) for any $N_v$. The 
result indeed agrees for $N_v = 2,3$
with the SUSY result obtained with U($1|1)$ \cite{dotv}:
\ba 
 \Sigma_\nu^\rmi{quenched}
 =  
 \mu \Bigl[ I_\nu(\mu) K_\nu(\mu) + I_{\nu+1}(\mu) K_{\nu-1}(\mu) \Bigr] + 
 \frac{\nu}{\mu}
 \;.  
 \la{def_Cqu}
 \label{zeromode}
\ea
For $\nu\neq 0$, the leading behaviour of this function at small 
and large mass is the same as
in the unquenched case at fixed topology, \eq\nr{Cnu}:
\be
 \Sigma_\nu^\rmi{quenched}(\mu) \approx 
 \left\{ \begin{array}{cc}
  {|\nu|}/{\mu}, & \mu \ll 1 \\
  1, & \mu \gg 1 
 \end{array}
 \right. \;.
\ee

%%%%%%%%%%%%%%%%%%%%%%%%%% SECTION %%%%%%%%%%%%%%%%%%%%%%%%%%%%%%%%%%%%%%%
\section{Conclusions}
\la{se:concl}

We have computed the two-point correlator of left-handed chiral charges,
as well as the three-point correlator of two left-handed charges and one 
strangeness violating weak operator~\cite{algo}, %% $O_{w}^{\pm}$, 
in SU(3) chiral 
perturbation theory in a finite volume and close to the chiral limit, 
at next-to-leading order in the $\epsilon$-expansion. The comparison 
of these observables with lattice data would in principle 
permit the extraction of the pion decay constant $F$, as well 
as the low-energy constant $g_{27}$,  
involved in the $\Delta I=3/2$ kaon decays
and in the kaon mixing parameter $\hat B_K$, with a minimal 
contamination from higher order corrections.
Whether this will be numerically feasible is still an open question. 

We have also performed the same calculations in the quenched theory, using
its replica formulation, and shown that these observables are only 
moderately affected. In particular,  the ratio defined 
in~\eq(\ref{norm_Cabcd}) is not only independent of the quark masses 
sufficiently close to the chiral limit, but also unaffected 
by quenching at this order. 
Obviously the low-energy constants obtained with the quenched 
theory nevertheless differ from those in full QCD.

%%%%%%%%%%%%%%%%%%%%%%% ACKNOWLEDGEMENTS %%%%%%%%%%%%%%%%%%%%%%%%%%%%%%%%%
\section*{Acknowledgements}

We are indebted to P.H.~Damgaard, 
L.~Giusti, C.~Hoelbling, K.~Jansen, L.~Lellouch, 
P.~Weisz, H.~Wittig and particularly M.~L\"uscher, for many 
useful discussions and suggestions. 

\newpage

%%%%%%%%%%%%%%%%%%%%%%%%%%%%% APPENDIX %%%%%%%%%%%%%%%%%%%%%%%%%%%%%%%%%%%

\appendix
\renewcommand{\thesection}{Appendix~\Alph{section}}
\renewcommand{\thesubsection}{\Alph{section}.\arabic{subsection}}
\renewcommand{\theequation}{\Alph{section}.\arabic{equation}}

%%%%%%%%%%%%%%%%%%%%%%%%%%%%%%%%%%%%%%%%%%%%%%%%%%%%%%%%%%%%%%%%%%%%%%%%

%\newpage

\section{SU(3) classification}
\la{app:su3}

For completeness, we reiterate in this Appendix some essential 
aspects of the SU(3) classification of four quark operators. 
We follow the tensor method discussed, e.g., in~\cite{hg2}. 

The tensors we need to consider are of the form $ O_{\a \b \c \d}$, 
symmetric under $(\a\leftrightarrow\b,
\c\leftrightarrow\d)$, and transforming under 
${\bf 3^*} \otimes {\bf 3^*} \otimes {\bf 3} \otimes {\bf 3}$ of SU(3). 
We then define the projected operators 
\ba
 {O}^\sigma_{\a\b\c\d} & \equiv &  
 (P_1^\sigma)_{\a\b\c\d;\ta\tb\tc\td}\;
 {O}_{\ta\tb\tc\td}, \la{O_sigma} \la{OP1} \\
 \hat {O}^\sigma_{\a\b\c\d} & \equiv &  
 (P_2^\sigma)_{\a\b\c\d;\ta\tb\tc\td}\;
 {O}^\sigma_{\ta\tb\tc\td},  \la{O_project} \la{OP2}
\ea
where $\sigma=\pm 1$. Here, 
with some redundancy in the symmetries of $P_1^\sigma$, 
\ba
 & & (P_1^\sigma)_{\a\b\c\d ; \ta\tb\tc\td} \equiv 
 \fr14 (\delta_{\a\ta} \delta_{\b\tb} + 
 \sigma \delta_{\a\tb}\delta_{\b\ta})
 (\delta_{\c\tc} \delta_{\d\td} + 
 \sigma \delta_{\c\td} \delta_{\d\tc}), \la{P1} \\
 & &  (P_2^\sigma)_{\a\b\c\d;\ta\tb\tc\td} \equiv  
 \delta_{\a\ta} \delta_{\b\tb} \delta_{\c\tc} \delta_{\d\td} 
 + \frac{1}{(N_f + 2\sigma)(N_f + \sigma)}
 ( 
 \delta_{\a\c} \delta_{\b\d} + \sigma
 \delta_{\a\d} \delta_{\b\c}) \delta_{\ta\tc} \delta_{\tb\td}
 \nn
 & & ~~~~~~~~ - \frac{1}{N_f + 2\sigma}
 (
 \delta_{\a\c} \delta_{\b\tb}  \delta_{\d\td} \delta_{\ta\tc}+
 \delta_{\b\d} \delta_{\a\ta} \delta_{\c\tc} \delta_{\tb\td}+
 \sigma \delta_{\a\d} \delta_{\b\tb} \delta_{\c\td} \delta_{\ta\tc}+
 \sigma \delta_{\b\c} \delta_{\a\ta} \delta_{\d\tc} \delta_{\tb\td}
), \hspace*{1.0cm} \la{P2}
\ea
where $N_f=3$.

It is easy to see that the antisymmetric tensor 
$\hat {\! O}^-_{\a \b \c \d} $
vanishes identically. The reason is that 
(as can be understood for instance by contracting 
with  $\epsilon_{\k\a \b} \epsilon_{\l\c \d}$)
it corresponds to a representation with dimension 8 just like  
$ R^-_{\l\k}$ defined in~\eq\nr{cRac}, 
but all such representations have already
been subtracted by the projection operator in~\eq\nr{P2}. 

Consequently, the reduction
of a general operator $O_{\a\b\c\d}$  proceeds as
\ba
  O_{\a\b\c\d} & = & 
 \hat {\! O}^+_{\a\b\c\d} +
 \sum_{\sigma = \pm 1}\biggl[ 
 \frac{1}{3(3+\sigma)}
 (\delta_{\a\c}\delta_{\b\d} + 
  \sigma \delta_{\a\d}\delta_{\b\c})  S^\sigma \nn 
  & & \hspace*{2cm} + \frac{1}{3 + 2\sigma} 
  (\delta_{\a\c}  R^\sigma_{\b\d} + 
  \delta_{\b\d}  R^\sigma_{\a\c} + 
  \sigma \delta_{\a\d}  R^\sigma_{\b\c} + \sigma \delta_{\b\c} 
   R^\sigma_{\a\d})
 \biggr],
\ea
where $\;\hat {\! O}^+_{\a\b\c\d}$ transforms
under the  representation with dimension 27, and
\ba
  S^\sigma & \equiv &  
 O^\sigma_{\k\l\k\l}, \\
  R^\sigma_{\a\c} & \equiv & 
  O^\sigma_{\a\k\c\k} - 
 \frac{1}{3} \delta_{\a\c}  S^\sigma. \la{cRac}
\ea
Here $ R^\pm_{\a\c}$'s have the dimension 8, 
while $ S^\sigma$ are singlets. 

Finally, let us note that in the chiral theory,
i.e.\ if we replace  
$ O_{\a\b\c\d} \to {[ {\cal O}_w ]}_{\a\b\c\d}$, 
\be
 {[  {\cal O}_w ]}_{\a\k\c\k} = 
 {[  {\cal O}_w ]}_{\k\b\k\d} = 0, \la{tless_su3}
\ee
so that
\be
 {[  {\cal O}_w ]}^\sigma_{\a\k\c\k} = 
 \frac{\sigma}{2} {[  {\cal O}_w ]}_{\a\k\k\c}, \quad
 {[  {\cal S}_w ]}^\sigma = \frac{\sigma}{2} 
 {[  {\cal O}_w ]}_{\l\k\k\l}, \quad
 {[  {\cal R}_w ]}^+_{\a\c} = 
 - {[  {\cal R}_w ]}^-_{\a\c}.
 \la{traceless_su3}
\ee

%%%%%%%%%%%%%%%%%%%%%%%%%% BIBLIOGRAPHY %%%%%%%%%%%%%%%%%%%%%%%%%%%%%%%%%%%
%%%%%%%%%%%%%%%%%%%%%%%%%%% REFERENCES %%%%%%%%%%%%%%%%%%%%%%%%%%%%%%%%%%%%

\end{document}